\documentclass[prd,twocolumn]{revtex4}

\usepackage{subfigure}
\usepackage{amssymb}
\usepackage{epsfig}
\usepackage{float}
\usepackage{caption}
\usepackage{color}
\usepackage{upgreek}
\usepackage{amsmath}

\newcommand{\be}{\begin{equation}}
\newcommand{\ee}{\end{equation}}
\newcommand{\bea}{\begin{eqnarray}}
\newcommand{\eea}{\end{eqnarray}}

\newcommand{\It}[1]{\mathit{#1}}
\newcommand{\Rm}[1]{\mathrm{#1}}

\begin{document}

\title{Relativistic Euler equations in cosmologies with non-linear structures}

\author{Christopher S. Gallagher}
\email{c.s.gallagher@qmul.ac.uk}
\affiliation{School of Physics and Astronomy, Queen Mary University of London, UK.}

\author{Timothy Clifton}
\email{t.clifton@qmul.ac.uk}
\affiliation{School of Physics and Astronomy, Queen Mary University of London, UK.}

\bibliographystyle{plain}

\begin{abstract}

We consider a new variant of cosmological perturbation theory that has been designed specifically to include non-linear density contrasts on scales $\lesssim 100$ Mpc, while still allowing for linear fluctuations on larger scales. This theory is used to derive the relativistic equations of Eulerian hydrodynamics in realistic cosmological scenarios that contain radiation and a cosmological constant, as well as matter that has been allowed to clump into galaxies and clusters of galaxies. These equations can be used to evolve energy densities and velocities in the presences of small-scale non-linear structures, and on scales all the way up to the horizon and beyond. The leading-order part of these equations reproduces the expected Newtonian equations, while subsequent orders prescribe relativistic corrections. We demonstrate that these evolution equations are consistent with maintaining the Einstein constraints, and hence that the system as a whole is mathematically well posed. The relativistic corrections that we derive are found to exhibit non-trivial interactions between perturbations on different scales, as well as the mixing of scalar, vector and tensor modes. They deviate from those that occur in both post-Friedmann and standard cosmological perturbation theory approaches, and point towards new relativistic effects that could be measurable by upcoming ultra-large-scale surveys.
\end{abstract}

%\pacs{98.80.Jk, 98.80.-k, 98.65.Dx, 04.25.Nx}

\maketitle

\section{Introduction}
\label{Introduction} 

The next generation of astronomical surveys will operate with a vastly increased scope, as compared to all predecessors \cite{SKA, euclid, LSST}. This exciting advancement opens up the possibility of extending tests of the non-linear aspects of Einstein's theory from isolated astrophysical systems into the cosmological regime \cite{ Kitching:2015fra, Laureijs:2011gra, Amendola:2016saw, Abell:2009aa}. Such tests will enable us to further constrain Einstein's equations in hitherto unprobed physical scenarios, and may even provide evidence of deviations away from general relativity \cite{ModGravCosmology:2011}. This is a topic of much contemporary interest, given the observed accelerating expansion of the Universe and the cosmological constant problems \cite{MaddoxSutherlandDE:1990, RiessDE:1998, PerlmutterDE:1998}.
 
Most approaches to modelling relativistic effects in cosmology are based on applying cosmological perturbation theory \cite{malik} to small fluctuations around an otherwise perfectly homogeneous and isotropic Robertson-Walker background (see e.g. \cite{Malik:2003mv, Huston:2009ac, Nakamura:2006rk}). Such studies are ideally suited to modelling the large-scale fluctuations in both gravitational and matter fields. They are not, however, well suited to studying the gravitational fields that result from the presence of structures with highly non-linear density contrasts. This fact arises because the density contrast itself is required to be small for the perturbative expansion to be well defined. Failure to meet this criterion means that terms that would normally occur at higher-orders are allowed to become large, and that the solutions to resulting equations are significantly altered. This difference is far from just a mathematical technicality, as can be seen clearly at even the Newtonian level, where additional non-linear terms in the Euler equations produce highly complicated phenomenology that is not readily reproduced by continuing perturbation theory to higher orders (see e.g. \cite{carlson}).

On the other hand, expansions of the gravitational field equations that are designed for use in the presence of highly non-linear structures are not always readily applicable to cosmology. The archetypal such scheme is the weak-field and slow-motion post-Newtonian expansion, which is very widely applied in the study of relativistic gravitational effects in the solar system and extra-solar binaries \cite{will,poisson}. This approach expands the retarded null Green's functions that arise from solving the full Einstein equations on spatial scales that are much smaller than the particle horizon, and as a result requires treating time derivatives acting on both matter and gravitational fields as being small compared to spatial derivatives acting on the same quantities. This seemingly small adjustment in the counting of powers of smallness means that non-linear density contrasts can be consistently included, and that the leading-order equations of motion for time-like particles include the required non-linear terms in the Euler equations. The higher-order equations are similarly applicable to situations that contain highly non-linear structures, and have been used to predict a host of relativistic gravitational effects in the solar system and beyond \cite{unreasonablepostnewtonian:2011, SelfForce:2010}.

Recent progress has been made in applying the techniques from post-Newtonian gravitational expansions into cosmology \cite{sanghai1, sanghai2, bruni0}, with numerical simulations even being performed to post-Newtonian orders of accuracy \cite{bruni,RDformalism, RDnature, RDgevolution, Adamek:2014xba}. These studies have yielded numerous interesting results, including the determination of the power spectra of higher-order gravitational potentials and the effects of small-scale gravitational fields on the large-scale expansion of space. Our current purpose is to combine these expansions, which are valid on scales $\lesssim 100$ Mpc, with the standard approach to cosmological perturbation theory, which is valid on scales $\gtrsim 100$ Mpc, where structures are largely still linear. We do this by simultaneously expanding all relevant equations in two parameters; one associated with the expansion parameter of cosmological perturbation theory ($\epsilon$), and the other associated with the order-of-smallness from post-Newtonian theory ($\eta$). Together, these two expansions cover the vast majority of structures that are expected to exist in the Universe (excluding regions of space-time in the immediate vicinity of black holes and neutron stars).

This two-parameter approach to cosmological perturbations was started in Ref. \cite{SRGTCKM}, and extended in Ref. \cite{Goldberg:2017} to include fluids with non-zero pressure and $\Lambda$. These studies considered the gauge issues involved in producing a well-defined two-parameter expansion in all required variables, identified the minimal sets of perturbed quantities required for a consistent study, and proceeded to calculate the perturbed field equations in terms of gauge-invariant variables. It was found that the large-scale cosmological perturbations are sourced by additional terms that are quadratic (or higher) in the quantities associated with the Newtonian fields, and that the post-Newtonian gravitational fields on small scales can be affected by the large-scale gravitational potentials in which they reside. These additional terms occur not only in the form of an effective fluid, but can also act to mix together the different scalar, vector and tensor degrees of freedom in the large-scale gravitational fields. Any effects of these terms should be considered to constitute new relativistic gravitational phenomena, due to the presence of small-scale non-linear structures.

In this paper, we present the corresponding relativistic version of the Eulerian equations of hydrodynamics in the two-parameter formalism and demonstrate that the Einstein constraint equations are consistently evolved by the full system of Einstein evolution and relativistic Euler equations. This demonstrates the system of equations proposed in Refs. \cite{SRGTCKM} and \cite{Goldberg:2017} are mathematically well posed, as well as providing a set of equations that can be used to determine the future properties of particles and fluids in such a setting. These equations will be of use for the accurate calculation of the matter power spectrum, as well as for calculating the growth rate of structure, and any other observables that are based on the positions of matter fields and that could hence be used to infer the effects of relativistic gravitational fields. In Section \ref{sec:2p} we outline the two-parameter formalism. Section \ref{sec:PerturbedConservationEquations} then contains a derivation of the relativistic Euler equations, with long equations being given in Appendix \ref{appendix}. For the busy reader, the important equations that result from Section \ref{sec:PerturbedConservationEquations} are all highlighted by the inclusion of boxes. Finally, in Section \ref{sec:conc} we end by discussing our results and concluding. 

We use Greek indices to represent space-time coordinates, and Latin indices for spatial coordinates. Dashes refer to differentiation with respect to conformal time, spatial derivatives are denoted by $\partial_i$, and $\nabla^2$ refers to the Laplacian operator associated with spatial partial derivatives in comoving coordinates. We use units in which $G = c = 1$ throughout, and choose to work in longitudinal gauge in both sectors of the perturbation theory.

\section{Two-parameter perturbation theory}
\label{sec:2p}

This section contains a summary of the two-parameter perturbation theory that was developed in Refs. \cite{SRGTCKM} and \cite{Goldberg:2017}, and which allows for higher-order relativistic gravitational effects to be calculated in the presence of non-linear structures. The first step in this formalism is to expand all quantities in the small parameters associated with each of the two expansions, such that any variable $Q$ can be written
\begin{align}
\label{q}
Q = \sum\limits_{n,m} \frac{1}{n!m!}Q^{(n,m)} \, ,
\end{align}
where $Q^{(n,m)}$ is a quantity of order $n$ in the cosmological perturbation theory expansion parameter $\epsilon$, and order $m$ in the post-Newtonian expansion parameter $\eta$. It is assumed that both $\epsilon \ll 1$ and $\eta \ll 1$. As is usual in post-Newtonian gravity, we treat time-derivatives of a quantity that has $m \neq 0$ as adding an extra order of smallness in that expansion, as compared to spatial derivatives. In cosmological perturbation theory, on the other hand, all derivatives are treated on an equal footing, and do not make a quantity smaller when it is perturbed only in that sector (i.e. if $m=0$). That is, for any quantity that is perturbed in the post-Newtonian expansion parameter $\eta$ we have
\be
{\rm P}_{,i} \sim \frac{{\rm P}}{L_N}
\qquad {\rm and} \qquad
{\rm P}^\prime \sim \eta \, \frac{{\rm P}}{L_N} \; ,
\ee
where $L_N$ is the typical length scale associated with that expansion, while for cosmological quantities we have
\be
{\rm C}_{,i} \sim \frac{{\rm C}}{L_C}
\qquad {\rm and} \qquad
{\rm C}^\prime \sim \frac{{\rm C}}{L_C} \; ,
\ee
where $L_C$ is the length scale associated with large-scale cosmological perturbations. For further details of post-Newtonian gravity the reader is referred to the textbooks by Will \cite{will} and Poisson \& Will \cite{poisson}, and for cosmological perturbation theory the review by Malik \& Wands \cite{malik}.

The framework developed in Refs. \cite{SRGTCKM} and \cite{Goldberg:2017} is constructed around a spatially flat Robertson-Walker geometry, 
\begin{align}
\label{flrw}
ds^2 = a^2(\tau) \left( \eta_{\mu \nu} + h_{\mu\nu} \right) \, ,
\end{align}
where $a(\tau)$ is the scale factor, $\tau$ is conformal time, and $h_{\mu \nu} \ll 1$ is a small perturbation to the conformal geometry, which we choose to write in longitudinal gauge. Cosmological perturbation theory is, of course, developed for use around just such a background. Post-Newtonian gravity, however, is usually developed around Minkowski space. Nevertheless, small adaptations allow it to be performed around a Robertson-Walker geometry in an entirely self-consistent and well-posed way, as long certain conditions are obeyed \cite{sanghai1,SRGTCKM}. All perturbations to the matter fields and metric are performed about this background, as prescribed in Eq. (\ref{flrw}), and then used to construct the field equations at each order in perturbations.

At leading order this gives
\begin{align} \label{newton1}
\mathcal{H}' &= -\frac{4\pi a^2}{3}(\bar{\uprho} + 3\bar{\Rm{p}}) +\frac{1}{3}\Lambda a^2\\[5pt]
\mathcal{H}^2 &= \frac{8\pi a^2}{3}\bar{\uprho} + \frac{1}{3}\Lambda a^2 \;, \label{newton2}
\end{align}
where $\mathcal{H}=a^{\prime}/a$ is the conformal Hubble rate, and where primes denote differentiation with respect to $\tau$. The quantities $\bar{\uprho}$ and $\bar{\Rm{p}}$ in these equations correspond to the spatial averages of the leading-order perturbations to the energy density and pressure, $\uprho^{(0,2)}$ and $\Rm{p}^{(0,0)}$. At the same order we obtain the equation governing the Newtonian gravitational potential:
\begin{align} \label{newton3}
\nabla^2 {\rm U} &= 4\pi a^2 \updelta \uprho_{\Rm{N}} \;,
\end{align}
where ${\rm U} = -\frac{1}{2}\big( h_{00}^{(0,2)} + h_{00}^{(1,1)}\big) = -\frac{1}{6}\big( h_{ii}^{(0,2)} + h_{ii}^{(1,1)}\big)$ plays the part of the Newtonian-level gravitational potential and $\updelta \uprho_{\Rm{N}} = \updelta \uprho^{(0,2)} + \uprho^{(1,1)}$ that of the Newtonian mass density. These equations are derived under the conditions that:

\begin{itemize}

\item[(i)] The ratio of the length scales associated with the post-Newtonian ($L_N$) and cosmological ($L_C$) perturbations is the same order of magnitude as the post-Newtonian expansion parameter itself, so that $L_N/L_C \sim \eta$. This means that
$$
{\rm C}^\prime \sim {\rm C}_{,i}  \sim {\rm P}^\prime \sim \eta \, {\rm P}_{,i}  \; ,
$$
so that spatial derivatives acting on post-Newtonian and mixed-order quantities are effectively made larger than other terms by $\eta^{-1}$. 

\item[(ii)] The magnitude of $\epsilon$ and $\eta$ are such that $\epsilon \sim 10^{-5}$ and $\eta^2 \sim 10^{-5}$. This choice is motivated by the structures that exist in the Universe, and the magnitude of the velocities associated with their components, as described in detail in Ref. \cite{Goldberg:2017}.

\item[(iii)] The integrated normal derivative of ${\rm U}$ over the closed boundary of every averaging domain in the Universe is always zero, such that $\int_{\partial \Omega} (n \cdot \partial {\rm U}) \, dA =0$. This is identically satisfied if periodic boundary conditions are chosen for spatial domains the size of the homogeneity scale.

\end{itemize}
Equations (\ref{newton1})-(\ref{newton3}) are exactly the equations that are expected from considering Newtonian gravitational fields on homogeneous and isotropically expanding background spaces, with the added bonus that it is the average of the Newtonian mass (along with background pressure and $\Lambda$) that drives the large-scale expansion.

\begin{widetext}

At next-to-leading order we find the scalar gravitational potentials obey the following equations:
\begin{align} \label{phieq}
\frac{1}{3} \nabla^2 \upphi + \mathcal{H}\upphi' + \mathcal{H}\uppsi' + \uppsi'' + 2\mathcal{H}'\upphi   
&= \frac{4\pi a^2}{3}(\updelta \uprho + \updelta \uprho_{\Rm{eff}} + 3 \updelta \Rm{p} + 3  \updelta \Rm{p}_{\Rm{eff}} ) + \frac{2}{3} (\mathcal{D}^{ij}\Rm{U}) \Rm{h}_{ij} - \frac{8 \pi a^2}{3} \updelta \uprho_{\Rm{N}} (\uppsi - \upphi) \;,
\\[5pt]
\frac{1}{3} \nabla^2 \uppsi - \mathcal{H}\uppsi' - \mathcal{H}^2\upphi  &= \frac{4\pi a^2}{3}(\updelta \uprho + \updelta \uprho_{\rm eff}) + \frac{1}{3} (\mathcal{D}^{ij}\Rm{U}) \Rm{h}_{ij} - \frac{16 \pi a^2}{3} \updelta {\uprho}_{\Rm{N}} \uppsi \; ,  \label{psieq}
\end{align}
where $\upphi = - \frac{1}{2}\big(  h_{00}^{(1,0)}+h_{00}^{(1,2)}+\frac{1}{2}h_{00}^{(0,4)}  \big)$ and $\uppsi = - \frac{1}{6}\big(  h_{ii}^{(1,0)}+h_{ii}^{(1,2)}+\frac{1}{2}h_{ii}^{(0,4)}  \big)$. On the right-hand side we have $\updelta \uprho = \uprho^{(1,0)} + \uprho^{(1,2)} + \frac{1}{2} \uprho^{(0,4)}$ and  $\updelta \Rm{p} = \Rm{p}^{(1,0)} + \Rm{p}^{(1,2)} + \frac{1}{2} \Rm{p}^{(0,4)}$, while the terms $\updelta \uprho_{\Rm{eff}}$ and $\updelta \Rm{p}_{\Rm{eff}}$ are effective fluid terms built from the solutions of the lower-order equations (\ref{newton1})-(\ref{newton3}). These effective fluid quantities are given below, in Eqs. (\ref{rhoeff})-(\ref{peff}). Finally, $\mathcal{D}_{ij}= \partial_i \partial_j - \frac{1}{3} \delta_{ij} \partial^k \partial_k$ denotes the trace-free second derivative operator on the conformal 3-space. 

The governing equation for the vector gravitational field can be written
\begin{align} \label{seq}
\nabla^2 \Rm{S}_i   + 4\partial_i \big( \uppsi'  + \mathcal{H}\upphi \big) +&16\pi a^2 \Big( (\bar{\uprho}+ \bar{\Rm{p}} + \updelta \uprho_{\Rm{N}})( \Rm{v}_{i} - \Rm{S}_i) + \Rm{Q}_i^{{\rm eff}} \Big)  = - 8\pi a^2 \updelta \uprho_{\Rm{N}} \Rm{S}_i - 2  \Rm{S}^j \partial_j \partial_i \Rm{U} \; ,
\end{align}
where $\Rm{S}_i = - \big( h_{0i}^{(0,3)} + h_{0i}^{(1,0)} + h_{0i}^{(1,2)} \big)$ and $\Rm{v}_i = v_i^{(1,0)}+v_i^{(0,2)}$. The quantity $\Rm{Q}_i^{{\rm eff}}$ is another effective fluid term, given in Eq. (\ref{qeff}). The largest part of this equation is particularly interesting as it is sourced entirely in terms of Newtonian-level quantities, and can hence be calculated using the output of standard cosmological $N$-body simulations \cite{bruni}. The explicit form of this equations is given by
\begin{align} \label{beq}
\nabla^2 \Rm{B}_i + 4 \partial_i (\Rm{U}_{\Rm{S}}' + \mathcal{H}\Rm{U}_{\Rm{S}}) + 16\pi a^2 (\bar{\uprho} + \bar{\Rm{p}} +\updelta \uprho_{\Rm{S}}) \Rm{v}_{\Rm{N}i}  =0 \;,
\end{align}
where $\Rm{B}_i = - h_{0i}^{(0,3)}$ is the leading post-Newtonian vector potential, $\Rm{U}_{\Rm{S}} = -\frac{1}{2}h_{00}^{(0,2)} =  -\frac{1}{6}h_{ii}^{(0,2)}$ is the Newtonian gravitational potential, $\updelta \uprho_{\Rm{S}} = \updelta \uprho^{(0,2)}$ is the inhomogeneous part of the Newtonian energy density, and $\Rm{v}_{\Rm{N}i}= v_i^{(0,1)}$ is the Newtonian peculiar velocity \footnote{The reason for defining these separate parts of $\Rm{U}$ and $\updelta \uprho_{\Rm{N}}$ will be made clear in the subsequent section on conservation equations.}. With the exception of $\nabla^2 \Rm{B}_i$, all other terms in this equation are Newtonian. 

The final equation, at the required order, is given by
\begin{align} \label{heq}
\mathcal{D}_{ij}(\upphi - \uppsi) - 2\mathcal{H}\partial_{(j}\Rm{S}_{i)}  - \partial_{(j}\Rm{S}'_{i)} +\nabla^2 \Rm{h}_{ij} - \Rm{h}''_{ij} - 2\mathcal{H} \Rm{h}'_{ij}
= -8\pi a^2 \Pi_{ij}^{\rm eff} - 8 \pi a^2 \updelta \uprho_{\Rm{N}} \Rm{h}_{ij} + 4 (\partial^k \partial_{\langle i}\Rm{U}) \Rm{h}_{j\rangle k} + 2(\mathcal{D}_{ij}\Rm{U}) (\upphi + \uppsi) \; ,
\end{align}
where $\Rm{h}_{ij} = \frac{1}{2} \big( h_{\langle ij \rangle}^{(1,0)} +  h_{\langle ij \rangle}^{(1,2)} +\frac{1}{2}  h_{\langle ij \rangle}^{(0,4)} \big) $, and where angular brackets denote the symmetric tracefree part of a quantity, so that $Q_{\langle i j \rangle} = Q_{(ij)} - \frac{1}{3} \delta_{ij} Q_{k k}$. The term $\Pi_{ij}^{\rm eff}$ is the final effective fluid term, and is given in Eq. (\ref{pieff}). This equation contains the terms that are usually decomposed to show that $\upphi=\uppsi$, and to gain separate governing equations for $\Rm{S}_i$ and $\Rm{h}_{ij}$. We have not performed this scalar-vector-tensor decomposition explicitly here as the coupling terms on the right-hand side complicate it somewhat, as does the different behaviour of the different types of terms under spatial derivatives. The lower-order version of this equation gives ${h}_{\langle ij \rangle}^{(0,2)} =  {h}_{\langle ij \rangle}^{(1,1)} = 0$. 

Equations (\ref{phieq})-(\ref{heq}) take the place of the usual gravitational field equations from cosmological perturbation theory, but with extra effective fluid terms generated from the non-linear structures in the Newtonian sector of the theory, and with additional mode-mixing terms on the right-hand side that couple the scalar and tensor modes as well as the scalar and vector modes. These terms collectively encode the relativistic gravitational effects of non-linear structures, and exist due to the non-linear nature of Einstein's equations. The effective fluid quantities that appear in Eqs. (\ref{phieq})-(\ref{heq}) are defined as follows:
\begin{align}
\updelta \uprho _{\rm eff} =&  (\bar{\uprho} + \bar{\Rm{p}} +\updelta \uprho_{\Rm{N}})\Rm{v}_{\Rm{N}}^2  
+ \frac{3}{4\pi a^2}\Big(\mathcal{H}^2 \Rm{U} + \mathcal{H}\Rm{U}' - \frac{1}{2}(\nabla \Rm{U} )^2  - \frac{4}{3}\Rm{U}\nabla^2\Rm{U} \Big) \;, \label{rhoeff} \\[5pt]
\updelta \Rm{p}_{\rm eff} =&  \frac{1}{3} (\bar{\uprho} + \bar{\Rm{p}} + \updelta \uprho_{\Rm{N}}) \Rm{v}_{\Rm{N}}^2 
- \frac{1}{4\pi a^2} \Big(\Rm{U}'' + 3\mathcal{H}\Rm{U}' + (2\mathcal{H}' + \mathcal{H}^2)\Rm{U} \Big) 
+  \frac{7}{24\pi a^2}(\nabla \Rm{U})^2  + \frac{1}{3\pi a^2} \Rm{U} \nabla^2 \Rm{U} \; , \label{peff} \\[5pt]
\Rm{Q}_i^{{\rm eff}} =& \updelta \uprho_{\Rm{N}} \Rm{v}_{\Rm{N}i} + \frac{1}{4\pi a^2} \partial_i \left(  \Rm{U}' + \mathcal{H} \Rm{U}   \right) \;,  \label{qeff} \\[5pt]
\Pi_{ij}^{\rm eff} =& (\bar{\uprho} + \bar{\Rm{p}} + \updelta \uprho_{\Rm{N}}) \Rm{v}_{\Rm{N}\langle i} \Rm{v}_{\Rm{N} j\rangle} 
- \frac{1}{4\pi a^2} \partial_{\langle i}\Rm{U} \partial_{j \rangle}\Rm{U}  - \frac{1}{2\pi a^2} \Rm{U} \mathcal{D}_{ij}\Rm{U} \;. \label{pieff} 
\end{align}
 Each of the effective fluid terms above can be constructed using the output of standard cosmological $N$-body simulations, and provide the potential for new physics on cosmological scales. The non-vanishing of the scalar part of $\Pi_{ij}^{\rm eff}$, for example, should be expected to produce a non-zero ``slip'' such that $\upphi \neq \uppsi$.

The reader may note that it is perfectly possible to define a two-parameter infinitesimal coordinate transformation, which in turn leads to the notion of two-parameter gauge transformations and two parameter gauge-invariant variables. This was done explicitly in Refs. \cite{SRGTCKM} and \cite{Goldberg:2017}, where it was shown that the quantities in the equations above can all be written in terms of gauge invariant quantities. For brevity, we omit the details of this from the current presentation. The reader may also note that not all possible perturbations are included in the equations above; for example, there is no term $\uprho^{(0,3)}$ or $h_{00}^{(0,3)}$. The inclusion of such terms does not facilitate the understanding of the relativistic corrections that we are interested in here, as they would obey exactly the same equations as the larger terms $\uprho^{(0,2)}$ and $h_{00}^{(0,2)}$. They are therefore not made explicit, but could be re-introduced if necessary at any point. The collection of variables given above is, in fact, the minimum set that can be consistently considered under general gauge transformations. That is, if any of the terms given above were artificially set to zero in one coordinate system, then the corresponding terms would be found to still exist in other coordinate systems, and so should be considered to exist in general.

\end{widetext}

\section{Conservation Equations}
\label{sec:PerturbedConservationEquations}

In this section we will present the stress-energy conservation equations for our two-parameter perturbation theory, for a spatially flat Robertson-Walker background geometry. This presentation will differ from the procedure used in standard cosmological perturbation theory, in which the linear-order conservation equations can be derived by straightforward manipulation of the linear-order field equations. Instead, we must take into account the fact that derivatives can change the size of objects they act upon in order to gain the correct equations. This complicates the situation considerably. The equations we derive in this section are only directly applicable to the single-stream case, but we expect multi-stream generalisations to follow straightforwardly.

In general, Einstein's equations contain four constraint equations and six evolution equations \cite{BaumgartenShapiro}. This number can be reduced in situations of high symmetry, such as in Friedmann-Lema\^{i}tre-Robertson-Walker space-times where there is one constraint equation ($\mathcal{H}^2 = \dots$) and one evolution equation ($\mathcal{H}' = \dots$). In this sense, one can identify Eqs. (\ref{newton2})-(\ref{newton3}) as constraint equations, and Eq. (\ref{newton1}) as an evolution equation. Likewise, at higher order, one can identify Eqs. (\ref{psieq}), (\ref{seq}) \& (\ref{beq}) as constraint equations, and Eqs. (\ref{phieq}) \& (\ref{heq}) as evolution equations. We can say that a set of constraint equations is maintained under evolution if after differentiating with respect to time, and substituting from the evolution equations, the same set of equations is recovered. This is an important property for a physical system to have, as it demonstrates that the system is neither overdetermined (a property that would result in different or additional constraint equations being generated at later times), nor underdetermined. In the analysis that follows will verify that the constraint equations from our two-parameter expansion are, in fact, consistently maintained under evolution. This will also serve as a check on the perturbed stress-energy conservation equations in this formalism, which can of course also be obtained from expanding $\nabla_{\mu} T^{\mu}_{\phantom{\mu} \nu} = 0$.

The derivation of our conservation equations will be presented in terms of the following gauge-invariant matter perturbations:
\begin{align} \label{matter}
\{ \updelta \uprho_{\rm S},  \updelta \uprho_{\rm M}, \updelta \uprho, \updelta \Rm{p} , \Rm{v}_{{\rm N}  {i}}, \Rm{v}_{{\rm M}  {i}} ,  \Rm{v}_{ {i}}  \} \, ,
\end{align}
where $\updelta \uprho_{\rm S} = \updelta \mathbf{\uprho}^{(0,2)}$ is the part of $\updelta \uprho_{\Rm N}$ that varies over short scales, and $\updelta \uprho_{\rm M} = \mathbf{\uprho}^{(1,1)}$ is the mixed part that is perturbed in both $\epsilon$ and $\eta$. This gives $\updelta \uprho_{\rm N}=\updelta \uprho_{\rm S}+\updelta \uprho_{\rm M}$ as the source term of Eq. (\ref{newton3}). Here we take the pressure term to be given by $\updelta \Rm{p} = \Rm{p}^{(1,0)} + \Rm{p}^{(1,2)} + \frac{1}{2} \Rm{p}^{(0,4)} + \dots$, where the ellipsis denotes higher-order terms that will be required in the manipulations that follow. Finally, we also introduce the mixed order velocity field $\Rm{v}_{{\rm M}  {i}} = v_i^{(1,1)}$. The other terms are all as defined above.

The corresponding set of gauge-invariant metric perturbations are given by
\begin{align} \label{gravity}
\{  {\rm U}_{\rm S}, {\rm U}_{\rm M}, \upphi, \uppsi,  {\rm B}_{ {i}}, {\rm A}_{ {i}},  {\rm h}_{ {ij}}  \} \, ,
\end{align}
where ${\rm U_S} = -\frac{1}{2} h_{00}^{(0,2)}$ and ${\rm U_M} = -\frac{1}{2} h_{00}^{(1,1)}$ are the short-wavelength and mixed-order parts of $U$, defined such that $\rm U = U_S +U_M$. We have also introduced the vector gravitational potential ${\rm A}_{ {i}} = - (h_{0i}^{(1,0)} + h_{0i}^{(1,2)})$, such that the full vector potential can be written $ {\rm S}_{ {i}}= {\rm B}_{ {i}}+ {\rm A}_{ {i}}$. Again, the other potentials are defined as in the previous section.

The new quantities introduced in Eqs. (\ref{matter}) and (\ref{gravity}) are motivated by close examination of the conservation equation, as presented below, after which it becomes clear that the evolution equations satisfied by $\updelta \uprho_{\rm S}$ and $\updelta \uprho_{\rm M}$ take different forms from one another. This motivates us to separate out the corresponding pairs of gravitational potentials  $\{  {\rm U}_{\rm S}, {\rm U}_{\rm M}  \}$ and $\{   {\rm B}_{ {i}},  {\rm A}_{ {i}}   \}$. Furthermore, the mixed order peculiar velocity, $\Rm{v}_{ {\rm M}  {i}}$, does not actually appear in any of the field equations (\ref{phieq})-(\ref{heq}), but is required to provide a complete set of closed conservation equations. Having introduced these variables for the purpose of performing calculations, where possible we will present the final Euler equations in terms of the original variables used in Eqs. (\ref{phieq})-(\ref{heq}).

\subsection{Conservation of the Friedmann and Newton-Poisson equations}

Let us start with the background Friedmann equations, to fix ideas. Taking the time derivative of the Friedmann equation (\ref{newton2}) gives
\begin{align}
\frac{\rm d}{\rm d \tau}(\mathcal{H}^2) = 2\mathcal{H}\mathcal{H}' = \frac{8\pi a^2}{3}\big(\bar{\uprho}' + 2\mathcal{H}\bar \uprho\big) + \frac{2}{3}\Lambda \mathcal{H}a^2 \;.
\end{align}
Now using the Raychaudhuri equation (\ref{newton1}) to eliminate $\mathcal{H}'$ gives
\begin{align}
2\mathcal{H}\mathcal{H}'  &= -\frac{8\pi a^2}{3}(\mathcal{H}\bar{\uprho} + 3\mathcal{H}\bar{\Rm{p}}) +\frac{2}{3}\Lambda \mathcal{H}a^2 \nonumber \; .
\end{align}
As usual, this equation shows that the Friedmann equation is recovered if and only if
\begin{align} \label{bgcon}
\boxed{\bar{\uprho}' + 3\mathcal{H}(\bar{\uprho}+\bar{\Rm{p}}) =0} \; ,
\end{align}
which can be straightforwardly verified to be the time component of the stress-energy conservation equation. This is exactly the same as the energy conservation equation from standard Friedmann cosmology, although here the background energy density $\bar{\uprho}$ should be understood as the average of the Newtonian mass, which is formally part of the perturbative expansion performed on small scales. This already shows an interesting link between the gravitational fields on large and small scales, which was exploited in Ref. \cite{sanghai3} to find consistency relations between super and sub-horizon gravitational potentials.

We can now repeat this procedure for the scalar gravitational potential $\rm U_S$. Differentiating the leading-order part of Eq. (\ref{newton3}) with respect to conformal time gives
\be
\nabla^2 \Rm{U_S'} = 4\pi a^2 (\updelta \uprho_{\Rm{S}}'+ 2\mathcal{H} \updelta \uprho_{\Rm{S}})  \; .    \nonumber 
\ee
It can be seen that taking the spatial derivative of equation (\ref{beq}) will result in another term $\nabla^2 \Rm{U_S}'$, which can be used to cancel the appearance of this term in the equation above. Explicitly, we obtain
\begin{multline}
\partial^i \nabla^2 \Rm{B}_i + 4 \nabla^2 (\Rm{U_S'} + \mathcal{H}\Rm{U_S}) \\+ 16\pi a^2 \partial^i\big((\bar{\uprho} + \bar{\Rm{p}} +\updelta \uprho_{\Rm{S}}) \Rm{v}_{\Rm{N} {i}}\big)  =0 \; , \nonumber 
\end{multline}
which on noting that $\partial^i \Rm{B}_i = 0$ gives
\begin{align}
\nabla^2 (\Rm{U_S'} + \mathcal{H}\Rm{U_S}) = -4\pi a^2 \partial^i\big((\bar{\uprho} + \bar{\Rm{p}} +\updelta \uprho_{\Rm{S}}) \Rm{v}_{\Rm{N} {i}}\big) \nonumber \; .
\end{align}
Cancelling $\nabla^2 \Rm{U_S}'$ from these equations then gives us back the constraint equation (\ref{newton3}), if and only if
\begin{align} \label{newtcon}
\boxed{ \updelta \uprho_{\Rm{S}}' + 3\mathcal{H} \updelta \uprho_{\Rm{S}} = -\partial^i\big((\bar{\uprho} + \bar{\rm p} +\updelta \uprho_{\Rm{S}}) \Rm{v}_{\Rm{N} {i}}\big) }\; ,
\end{align} 
which is the Newtonian equation for the conservation of energy on an expanding background. The reader may note that taking the spatial divergence of the vector equation (\ref{beq}) does not change the relative size of any terms, as all quantities are either post-Newtonian or mixed in perturbations (i.e. have $n\neq 0$, from Eq. (\ref{q})). Both equations (\ref{bgcon}) and (\ref{newtcon}) involve quantities of order $\mathcal{O}(\eta^3/L_N^3)$, indicating that the non-linear leading-order Newtonian fluctuations in the mass density can be of the same size as (or larger than) their mean values. This pleasing feature allows the construction of models where density contrasts on small scales can be very large.

At all subsequent orders, where terms from the post-Newtonian and cosmological sectors of the theory appear concurrently, we must be more careful, as post-Newtonian quantities become smaller by factors of $\eta$ under the action of a time derivative. The result of this is that certain terms are promoted to lower-order by spatial differentiation. This will be very important in obtaining the Euler equations at higher-orders in our expansion: A naive derivation of the same equations, by differentiating the field equations from Section \ref{sec:2p} only, would result in errors.

\begin{widetext}

Let us now consider the time derivative of the vector equation (\ref{beq}), which does not contain any terms that change size under the action of a spatial derivative, as it contains post-Newtonian terms only. This gives
\begin{align} \label{vector4dash}
\nabla^2 \Rm{B}_i' + 4 \partial_i (\Rm{U''_S}+ \mathcal{H}'\Rm{U_S} + \mathcal{H}\Rm{U'_S}) + 16\pi a^2 \big((\bar{\uprho} + \bar{\Rm{p}} +\updelta \uprho_{\Rm{S}}) \Rm{v}_{\Rm{N}i}\big)' + 16\pi a^2 \big(2\mathcal{H}(\bar{\uprho} + \bar{\Rm{p}} +\updelta \uprho_{\Rm{S}}) \Rm{v}_{\Rm{N}i}\big)  =0 \;. 
\end{align}
Likewise, the first non-trivial order of the trace-free field equation (\ref{heq}) is
\begin{multline}
\mathcal{D}_{ij}(\upphi - \uppsi) - 2\mathcal{H}\partial_{(j}\Rm{A}_{i)}  - \partial_{(j}\Rm{A}'_{i)} - 2\mathcal{H}\partial_{(j}\Rm{B}_{i)}  - \partial_{(j}\Rm{B}'_{i)} +\nabla^2\Rm{h}_{ij} - \Rm{h}''_{ij} - 2\mathcal{H}\Rm{h}'_{ij} \\ 
= -8\pi a^2 \Pi_{ij}^{\rm eff} - 8 \pi a^2 \updelta \uprho_{\Rm{S}} \Rm{h}_{ij} + 4 (\partial^k \partial_{\langle i}\Rm{U_S}) \Rm{h}_{j\rangle k} + 2(\mathcal{D}_{ij}\Rm{U_S}) (\upphi + \uppsi)\;.
\end{multline}
Take the leading-order part of the divergence of this equation, and using the result 
\begin{align}
\partial^k \partial^j \big(\partial_{\langle i}\Rm{U_S}\;\Rm{h}_{j \rangle k} \big)= \frac{1}{6} \partial_{ i}\mathcal{D}^{jk}\Rm{U_S}\Rm{h}_{j k} + 2\pi a^2 (\partial^k \updelta \uprho_{\Rm{S}})\Rm{h}_{ik} -  \frac{1}{3}\updelta_{ij} \partial^j( \mathcal{D}^{lk}\Rm{U_S})\Rm{h}_{lk}\;,
\end{align}
we obtain
\begin{align} \label{divtracefree4}
\frac{2}{3}\nabla^2 \partial_i (\upphi - \uppsi)-\frac{1}{2}\nabla^2(\Rm{B}'_i +\mathcal{H}\Rm{B}_i) =& -8\pi a^2 \partial^j \Pi_{ij}^{\rm eff} + \frac{2}{3}(\partial_i \mathcal{D}^{kj}\Rm{U_S})\Rm{h}_{kj} + \frac{16\pi a^2}{3}(\partial_i \updelta \uprho_{\Rm{S}})(\upphi + \uppsi )\;. 
\end{align}
We can proceed further by looking at the leading-order parts of the spatial gradients of the scalar gravitational field equations (\ref{phieq})-(\ref{psieq}). These can be combined to obtain
\begin{align} \label{scalargrad4}
4\pi a^2( \partial_i \updelta \Rm{p} + \partial_i \updelta \Rm{p}_{\rm eff} ) + \frac{1}{3} (\partial_i \mathcal{D}^{kj}\Rm{U_S}) \Rm{h}_{kj} +  \frac{8\pi a^2}{3}(\partial_i \updelta \uprho_{\Rm{S}})(\upphi + \uppsi) = \frac{1}{3}\nabla^2 \partial_i (\upphi - \uppsi)\;.
\end{align}
Substituting (\ref{scalargrad4}) into (\ref{divtracefree4}), and using Eq. (\ref{beq}), then yields the following expression for $\nabla^2 \Rm{B}'_i$:
\begin{align}
\nabla^2 \Rm{B}'_i = 16\pi a^2 (\partial_i \updelta \Rm{p} + \partial^j \Pi_{ij}^{\rm eff} + \partial_i \delta \Rm{p}_{\rm eff}) + 4\partial_i(2\mathcal{H}\Rm{U'_S} + 2\mathcal{H}^2\Rm{U_S}) + 16\pi a^2\Big(2\mathcal{H} \big( \Rm{v}_{\Rm{N}i}(\bar{\uprho} + \bar{\Rm{p}} + \updelta \uprho_{\Rm{S}})\big)\Big)\;. 
\end{align}
Now, substituting this into Eq. (\ref{vector4dash}) we obtain
\begin{multline} \label{mid3}
0 = 4\partial_i\big(\Rm{U''_S} + 3\mathcal{H}\Rm{U'_S} + (2\mathcal{H}^2 + \mathcal{H'})\Rm{U_S}\big) + 16\pi a^2 \big( \partial^j \Pi_{ij}^{\rm eff} + \partial_i \updelta \Rm{p} + \partial_i \updelta \Rm{p}_{\rm eff}\big)  \\
+16\pi a^2\Big( \big(\Rm{v}_{\Rm{N}i}(\bar{\uprho} + \bar{ {\rm p}} + \updelta \uprho_{\Rm{S}}) \big)' + 4\mathcal{H}\big(\Rm{v}_{\Rm{N}i}(\bar{\uprho} + \bar{\Rm{p}} + \updelta \uprho_{\Rm{S}}) \big)\Big)\;. 
\end{multline}
Next, we can use the following relation derived from taking spatial derivatives of the effective fluid quantities:
\begin{align}
16\pi a^2 (\partial^j \Pi_{ij}^{\rm eff} + \partial_i \updelta \Rm{p}_{\rm eff}) = 16\pi a^2 \partial^j\Big( \Rm{v}_{\Rm{N}i}\Rm{v}_{\Rm{N}j}(\bar{\uprho} + \bar{\Rm{p}} + \updelta \uprho_{\Rm{S}})\Big) 
 - 4\partial_i\big(\Rm{U''_S} + 3\mathcal{H}\Rm{U'_S} + (2\mathcal{H}' + \mathcal{H}^2)\Rm{U_S}\big)\;.
\end{align}
This, in conjunction with the spatially averaged Newtonian field equations (\ref{newton1}) and (\ref{newton2}), allows Eq. (\ref{mid3}) to be written as
\begin{align}
 \big(\Rm{v}_{\Rm{N}i}(\bar{\uprho} + \bar{\Rm{p}} + \updelta \uprho_{\Rm{S}}) \big)' + \partial^j\big( \Rm{v}_{\Rm{N}i}\Rm{v}_{\Rm{N}j}(\bar{\uprho} + \bar{\Rm{p}} + \updelta \uprho_{\Rm{S}})\big) +  \partial_i \updelta \Rm{p} 
+  4\mathcal{H}\big(\Rm{v}_{\Rm{N}i}(\bar{\uprho} + \bar{\Rm{p}} + \updelta \uprho_{\Rm{S}}) \big)  + (\partial_j  {\rm U_S}) (\bar{\uprho} + \bar{\rm p} + \updelta \uprho_{\rm S})  =0\;.
\end{align}
Further simplification can be made using the Newtonian-level energy conservation equations (\ref{bgcon}) and (\ref{newtcon}) to obtain the more familiar form.
\begin{align} \label{newteul}
\boxed{ {\rm v}_{ {\rm N} {j}}'(\bar{\uprho} + \bar{\rm p} + \updelta \uprho_{\rm S})  +  {\rm v}_{ {\rm N} {i}}\partial^{ {i}}   {\rm v}_{ {\rm N}  {j}}(\bar{\uprho} + \bar{\rm p} + \updelta \uprho_{\rm S})  +  {\rm v}_{ {\rm N} {j}}\bar{\rm p}' +  {\rm v}_{ {\rm N} {j}}\mathcal{H}(\bar{\uprho} + \bar{\rm p}+ \updelta \uprho_{\rm S})  = - (\partial_j  {\rm U_S}) (\bar{\uprho} + \bar{\rm p} + \updelta \uprho_{\rm S})   - \partial_{ {j}} (\updelta  {\rm p}) } \;.
\end{align}
 We have checked that this is almost precisely the generalisation of the Euler equation that is derived by direct calculation of the spatial components of the stress-energy conservation equations up to terms $\mathcal{O}(\eta^4/L_N^3)$, as expected from standard post-Newtonian theory on an expanding background. The only addition here is a time-dependent background pressure that would be considered negligible in typical post-Newtonian systems \cite{will}, but could be relevant in cosmological systems \cite{Peebles}.

Finally, we can obtain the conservation equation for $\updelta \uprho_{\Rm{M}}$ by taking the next-to-leading-order part of the spatial divergence of the vector gravitational field equation (\ref{seq}). This gives
\begin{align}
16\pi a^2 (\partial^i  \updelta \uprho_{\Rm{S}})( \Rm{v}_{i} - \Rm{A}_i) + 16 \pi a^2 \partial^i \Rm{v}_{\Rm{P}i}(\bar{\uprho} + \bar{\Rm{p}} + \updelta \uprho_{\Rm{S}})  + 16\pi a^2 \partial^i  \Rm{Q}_i^{{\rm eff}} = - \partial^i (8\pi a^2 \updelta \uprho_{\Rm{S}}) \Rm{A}_i - 2 \Rm{A}^j \partial_j \nabla^2 \Rm{U_S} \;,
\end{align}
 where $\Rm{v}_{\Rm{P}i} = v_i^{(0,2)}$ is the first post-Newtonian correction to the peculiar velocity. On using the leading-order part of the Newton-Poisson equation (\ref{newton3}), simplifies down to
$
\partial^i  \Rm{Q}_i^{{\rm eff}}  = - (\partial^i  \updelta \uprho_{\Rm{S}}) \Rm{v}_{i}
$. Using the definition of $\Rm{Q}_i^{{\rm eff}}$ from Eq. (\ref{qeff}), and the next-to-leading-order part of Eq. (\ref{newton3}), we then find
\begin{align} \label{mixcon}
\boxed{
\updelta \uprho_{\Rm{M}}' + 3\mathcal{H} \updelta \uprho_{\Rm{M}} = - \partial^i \Rm{v}_{\Rm{P}i}(\bar{\uprho} + \bar{\Rm{p}}+ \updelta \uprho_{\Rm{S}})   - (\partial^i  \updelta \uprho_{\Rm{S}}) \Rm{v}_{i} - \partial^i(\updelta \uprho_{\Rm{M}} \Rm{v}_{\Rm{N}i}) }\; . 
\end{align}
This is the first conservation equation we have found that explicitly links post-Newtonian, mixed, and cosmological quantities. It has no analogue in either post-Newtonian gravity or cosmological perturbation theory, as it involves terms from both such expansion schemes, and is therefore the first term to describe the effect of the interactions between these two sectors on the evolution of the energy density. We have again verified that this equation can be directly obtained by expanding the time component of the stress-energy conservation equation up to terms $\mathcal{O}(\eta^4/L_N^3)$.

The evolution equation for $\updelta \uprho_{\Rm N}$ can be obtained by combining Eqs. (\ref{newtcon}) and (\ref{mixcon}), to get
\begin{align} \label{enfin}
\updelta \uprho_{\Rm{N}}' + 3\mathcal{H} \updelta \uprho_{\Rm{N}} = -\partial^i\big((\bar{\uprho} + \bar{\rm p} +\updelta \uprho_{\Rm{N}}) (\Rm{v}_{\Rm{N} {i}} + \Rm{v}_{i})\big) + \mathcal{O}(\eta^5) \; ,
\end{align} 
where we have used $\updelta \uprho_{\Rm{N}} = \updelta \uprho_{\Rm{S}} + \updelta \uprho_{\Rm{M}}$. Taking the leading-order part of this equation recovers Eq. (\ref{newtcon}), while taking the next-to-leading order gives Eq. (\ref{mixcon}). 

\subsection{Conservation of the cosmological perturbation equations}

Having verified that the constraints are consistently evolved for the background and Newtonian sectors of the field equations, we now wish to perform the corresponding calculation for the cosmological perturbation equations (\ref{phieq})-(\ref{heq}). This requires considering terms up to $\mathcal{O}(\eta^5/L_N^3)$. The explicit calculations involved in performing this analysis are somewhat lengthy, and are therefore detailed in Appendix \ref{appendix}. Here we will present the results only, in the form of the relevant Euler equations.

The resultant energy conservation equation for $\updelta \uprho$ is
\begin{equation}\label{rhoCcontinuity}
\boxed{
\begin{aligned} 
&\updelta \uprho' + 3\mathcal{H}(\updelta \uprho +  \updelta \Rm{p}) - 3(\uppsi'+ \rm U^{\prime}_{\rm S})(\bar{\uprho} + \bar{ {\rm p}} + \updelta \uprho_{ {\rm S}} ) = - \partial^{\It{i}}\big( {\rm v}_{ {\rm M} \It{i}}(\bar{\uprho} + \bar{ {\rm p}} +\updelta \uprho_{\rm S})\big)   - \partial^{\It{i}} {\rm v}_{ \Rm{C} \It{i}}(\bar{\uprho} + \bar{ {\rm p}} + \updelta \uprho_{ {\rm S}} ) 
\\
& - \partial^{ \It{i}}\big({\rm v}_{ {\rm N} \It{i}}(\updelta \uprho + \updelta  {\rm p})\big)  -  {\rm v}_{ \It{i}}\partial^{ \It{i}} \updelta \uprho_{ {\rm M}} - (\partial^i \Rm{v}_{\Rm{P}i}) \updelta \uprho_{ {\rm M}} - \partial^{ \It{i}} \big( {\rm v}_{ {\rm N} {i}}(\bar{\uprho} + \bar{ {\rm p}} + \updelta \uprho_{ {\rm S}} )\big)(\upphi+ \rm U_{\rm S})  \\
&+ \partial^{ \It{i}}{\rm U_S}(\bar{\uprho}+\bar{ {\rm p}}+\updelta \uprho_{ {\rm S}}){\rm v}_{ {\rm N} \It{i}} 
 - \big(  {\rm v}_{ {\rm N}}^2(\bar{\uprho} + \bar{\rm p} + \updelta \uprho_{ {\rm S}} )\big)' - 4 \mathcal{H} {\rm v}_{ {\rm N}}^2(\bar{\uprho} + \bar{ {\rm p}} + \updelta \uprho_{ {\rm S}} )  - \frac{1}{2} \:\partial_i ( \rm{v}_{\rm{N}}^2 \:\rm{v}_{\rm{N}}^{\it{i}}(\bar{\uprho} + \bar{\rm{p}} +\updelta\uprho_{\rm{S}}) ) \; , 
\end{aligned}
}
\end{equation}
where $\Rm{v}_{\Rm{C} i} = v_i^{(1,0)}$. It can be seen that this equation is sourced by typical cosmological perturbation theory terms, such as $ 3\uppsi'(\bar{\uprho} + \bar{ {\rm p}})$, but also byproducts of leading-order terms such as $\mathcal{H} {\rm v}_{ {\rm N}}^2(\bar{\uprho} + \bar{ {\rm p}} + \updelta \uprho_{ {\rm S}} )$ and mixed-order terms like $ \partial^{ {i}}\big((\bar{\uprho} + \bar{ {\rm p}} +\updelta \uprho_{\rm S}) {\rm v}_{ {\rm M} {i}}\big)$. Some care is required in interpreting this equation as $\updelta \uprho$ has a different meaning when it appears under a spatial gradient, as the cosmological contribution $ \uprho^{(1,0)}$ gains an extra relative order-of-smallness under the action of a gradient compared to the mixed and post-Newtonian contributions $\uprho^{(1,2)}$ and $\uprho^{(0,4)}$. Thus, when considering the term $\partial^{ {i}}\big({\rm v}_{ {\rm N} {i}}(\updelta \uprho + \updelta  {\rm p})\big)$ the reader should understand the product term ${\rm v}_{ {\rm N} {i}}\partial^i(\updelta \uprho + \updelta  {\rm p})$  to include only mixed and post-Newtonian contributions,  whilst the product term $(\partial^{ {i}}{\rm v}_{ {\rm N} {i}})(\updelta \uprho + \updelta  {\rm p})$ should be understood to include all contributions, since the spatial gradient acting on the Newtonian peculiar velocity does not alter its size. The term $\updelta \uprho' $ should be understood to include all contributions, as the action of a conformal time derivative does not make any term small or larger than any other, regardless of their origin.

Likewise, the Euler equation for the velocity field ${\rm v}_{ {i}}$ is found to be
\begin{equation} \label{vCeuler}
\boxed{
\begin{aligned}
&\big(( {\rm v}_{ \It{i}}- {\rm A}_{ \It{i}})(\bar{\uprho}+ \bar{ {\rm p}} + \updelta \uprho_{ {\rm S}}) + {\rm v}_ {\rm N \It{i}} \updelta \uprho_{\rm M} \big)' + 4 \mathcal{H}\big(( {\rm v}_{ \It{i}}- {\rm A}_{ \It{i}})(\bar{\uprho}+ \bar{ {\rm p}} + \updelta \uprho_{ {\rm S}}) + {\rm v}_ {\rm N \It{i}}\updelta \uprho_{\rm M} \big) 
+ ({\rm v}_{ \It{i}} -{\rm A}_{\It{i}})\partial^{\It{j}} \big({\rm v}_{\rm N \It{j}}(\bar{\uprho}+ \bar{ {\rm p}} + \updelta \uprho_{ {\rm S}}) \big) 
 \\
& + {\rm v}_{ \It{j}} \partial^{\It{j}} \big({\rm v}_{\rm N \It{i}}(\bar{\uprho}+ \bar{ {\rm p}} + \updelta \uprho_{ {\rm S}}) \big) + (\partial^j \Rm{v}_{\Rm{P} i})\Rm{v}_{\Rm{N}j}(\bar{\uprho} + \bar{\Rm{p}} + \updelta \uprho_{\Rm{S}}) +  (\partial^j \Rm{v}_{\Rm{P} j})\Rm{v}_{\Rm{N}i}(\bar{\uprho} + \bar{\Rm{p}} + \updelta \uprho_{\Rm{S}})   +   \partial^{ \It{j}} ( {\rm v}_{ {\rm N} {i}} {\rm v}_{ {\rm N} \It{j}} \updelta \uprho_{ {\rm M}}) \\ \nonumber
& =-  \partial_{ \It{i}} (\upphi + { {\rm U_M}})(\bar{\uprho}+ \bar{ {\rm p}} + \updelta \uprho_{ {\rm S}}) - \updelta \uprho_{\rm M} \partial_{\It{i}}  \rm U_{\rm S} - \partial_{\It{i}}(\updelta  {\rm p} )   \; . 
\end{aligned}
}
\end{equation}
This is clearly a vector equation, and the evolution of the irrotational part of ${\rm v}_{ {i}}$ and vector gravitational potential ${\rm A}_{ {i}}$ can be seen to be given by its divergence and divergence-less parts, respectively. As before, we choose not to do this decomposition explicitly here, as the product terms and the rules associated with derivatives acting on different types of fields will complicate the results. This equation is reminiscent of the corresponding Euler equation from standard cosmological perturbation theory, with extra terms due to the existence of the non-linear structures on small scales. We note in particular that the mixed term $\partial^{ {j}} ( {\rm v}_{ {\rm N} {i}} {\rm v}_{ {\rm N} {j}} \updelta \uprho_{ {\rm M}})$ acts as a source for cosmological peculiar velocities and vector gravitational perturbations. We have verified that both Eq. (\ref{rhoCcontinuity}) and Eq. (\ref{vCeuler}) are recovered from the stress-energy conservation equations at $\mathcal{O}(\eta^5/L_N^3)$.

We can combine Eqs.  (\ref{newteul}) and (\ref{vCeuler}) to write a multi-order Euler equation for the evolution of the Newtonian and cosmological peculiar velocities;
\begin{align}\label{combinedeuler}
& \big((\Rm{v}_{\Rm{N}i}+\Rm{v}_{i}-\Rm{S}_i)(\bar{\uprho} + \bar{\Rm{p}} + \updelta \uprho_{\Rm{N}}) \big)'  +  4\mathcal{H}\big((\Rm{v}_{\Rm{N}i}+\Rm{v}_{i}-\Rm{S}_i)(\bar{\uprho} + \bar{\Rm{p}} + \updelta \uprho_{\Rm{N}}) \big) + \partial_j  ({\rm U} + \upphi) (\bar{\uprho} + \bar{\rm p} + \updelta \uprho_{\rm N}) \nonumber \\
 &+ \partial^j\big((\Rm{v}_{\Rm{N}i}+\Rm{v}_{i}-\Rm{S}_i)(\Rm{v}_{\Rm{N}j} +\Rm{v}_{j})(\bar{\uprho} + \bar{\Rm{p}} + \updelta \uprho_{\Rm{N}})\big) +  \partial_i \updelta \Rm{p}   =0 + \mathcal{O}(\eta^6)\;.
\end{align}
The evolution equation for $\Rm{v}_{\Rm{N} {i}}$ given in Eq. (\ref{newteul}) is then recovered by taking the order $\mathcal{O}(\eta^4)$ part of this equation, while Eq. (\ref{vCeuler}) is recovered by taking the order $\big(\mathcal{O}(\eta^5) \big)$ part.

The final equation required to complete a closed set of evolution equations for the set $\{ \updelta \uprho_{\rm S}, \updelta \uprho_{\rm M}, \updelta \uprho, \delta {\rm p}, \Rm{v}_{ {\rm N}  {i}}, \Rm{v}_{ {\rm M}  {i}} ,  \Rm{v}_{ {i}} \} $ is an evolution equation for $\Rm{v}_{\Rm{M}i}$. This is obtained from the perturbed stress-energy conservation equations at $\mathcal{O}(\eta^6/L_N^3)$, and is as follows:  
\begin{equation} \label{vmeuler}
\boxed{
\begin{aligned}
&\big( \Rm{v}_{\Rm{N}i} (\updelta \uprho + \updelta \Rm{p}) + (\Rm{v}_{\Rm{M}i} - \Rm{B}_i)(\bar{\uprho} + \bar{\Rm{p}} + \updelta \uprho_{\Rm{S}}) + (\Rm{v}_{i} - \Rm{A}_i)\updelta \uprho_{\Rm{M}} + 2\Rm{h}_{ij} \Rm{v}_{\Rm{N}}^j (\bar{\uprho} + \bar{\Rm{p}} + \updelta \uprho_{\Rm{S}})\big)' 
 \\
&+4\mathcal{H}\big(\Rm{v}_{\Rm{N}i} (\updelta \uprho + \updelta \Rm{p}) + (\Rm{v}_{\Rm{M}i} - \Rm{B}_i)(\bar{\uprho} + \bar{\Rm{p}} + \updelta \uprho_{\Rm{S}})  + (\Rm{v}_{i} - \Rm{A}_i) \updelta \uprho_{\Rm{M}}  + 2\Rm{h}_{ij} \Rm{v}_{\Rm{N}}^j (\bar{\uprho} + \bar{\Rm{p}} + \updelta \uprho_{\Rm{S}})\big) 
\\
&- \Big( \big( \Rm{v}_{\Rm{N}i}(\bar{\uprho} + \bar{\Rm{p}} + \updelta \uprho_{\Rm{S}})\big)' + 4\mathcal{H} \Rm{v}_{\Rm{N}i}(\bar{\uprho} + \bar{\Rm{p}} + \updelta \uprho_{\Rm{S}}) \Big) (\upphi + 2\uppsi + 3\Rm{U_S})  - 5 \Rm{v}_{\Rm{N}i}(\bar{\uprho} + \bar{\Rm{p}} + \updelta \uprho_{\Rm{S}})(\uppsi' + \Rm{U'_S}) 
\\
&+ \frac{1}{2}\big(\Rm{v}_{\Rm{N}}^2 \Rm{v}_{\Rm{N}i} (\bar{\uprho} + \bar{\Rm{p}} + \updelta \uprho_{\Rm{S}}) \big) ' + 2 \mathcal{H}(\Rm{v}_{\Rm{N}}^2 \Rm{v}_{\Rm{N}i} (\bar{\uprho} + \bar{\Rm{p}} + \updelta \uprho_{\Rm{S}}) \big) =
\\ 
&- \partial^j \big( \Rm{v}_{\Rm{N}j}(\Rm{v}_{\Rm{M}i} - \Rm{B}_i)(\bar{\uprho} + \bar{\Rm{p}} + \updelta \uprho_{\Rm{S}})\big) - \partial^j \big( \Rm{v}_{\Rm{N}i}\Rm{v}_{\Rm{M}j}(\bar{\uprho} + \bar{\Rm{p}} + \updelta \uprho_{\Rm{S}}) \big) - ({\rm v_{\It{i}}} - {\rm A_{\It{i}}}) {\rm v_{\It{j}}} \partial^{\It{j}} \updelta \uprho_{\rm S} 
\\ 
&- ({\rm v_{\It{i}}} - {\rm A_{\It{i}}})\partial^{\It{j}} {\rm v_{\Rm{P}\It{j}}} (\bar{\uprho} + \bar{\Rm{p}} + \updelta \uprho_{\Rm{S}})   - {\rm v_{\It{j}}} \partial^{\It{j}} {\rm v_{\Rm{P}\It{i}}} (\bar{\uprho} + \bar{\Rm{p}} + \updelta \uprho_{\Rm{S}})   - \partial^j \big(\Rm{v}_{\Rm{N}i}\Rm{v}_{\Rm{N}j}(\updelta \uprho + \updelta {\rm p}) \big) 
 \\
& -  \Rm{v}_{\Rm{N}j}\partial^j(\Rm{v}_{\Rm{C}i} - \Rm{A}_i) (\bar{\uprho} + \bar{\Rm{p}} + \updelta \uprho_{\Rm{S}}) - \Rm{v}_{\Rm{N}i}\partial^j \Rm{v}_{\Rm{C}j} (\bar{\uprho} + \bar{\Rm{p}} + \updelta \uprho_{\Rm{S}}) -  ({\rm v_{\It{i}}} - {\rm A_{\It{i}}}) \partial^{\It{j}} (\updelta \uprho_{\rm M} \Rm{v}_{\Rm{N}j}  ) - (\partial^{\It{j}} {\rm v_{\Rm{P}\It{i}}}) \updelta \uprho_{\rm M} \Rm{v}_{\Rm{N}j}   
\\
&-  {\rm v_{\It{j}}} \partial^{\It{j}} (\updelta \uprho_{\rm M} \Rm{v}_{\Rm{N}i}  )   - (\partial^{\It{j}} {\rm v_{\Rm{P}\It{j}}})  \updelta \uprho_{\rm M} \Rm{v}_{\Rm{N}i}   - 2\Rm{h}_{ij}\Rm{v}_{\Rm{N}}^j \partial^k \big( \Rm{v}_{\Rm{N}k}(\bar{\uprho} + \bar{\Rm{p}} + \updelta \uprho_{\Rm{S}})\big) - 2\Rm{h}_{ik}\Rm{v}_{\Rm{N}}^j \partial^k \Rm{v}_{\Rm{N}j}(\bar{\uprho} + \bar{\Rm{p}} + \updelta \uprho_{\Rm{S}})
 \\
& + 4 \Rm{v}_{\Rm{N}j}\Rm{v}_{\Rm{N}i} (\bar{\uprho} + \bar{\Rm{p}} + \updelta \uprho_{\Rm{S}}) \partial^j \Rm{U}_{\Rm{S}}  + 2( \Rm{U}_{\Rm{S}} + \uppsi) \partial^j \big(\Rm{v}_{\Rm{N}j}\Rm{v}_{\Rm{N}i} (\bar{\uprho} + \bar{\Rm{p}} + \updelta \uprho_{\Rm{S}})\big) -  2 \Rm{v}_{\Rm{N}}^2 (\bar{\uprho} + \bar{\Rm{p}} + \updelta \uprho_{\Rm{S}}) \partial_i \Rm{U}_{\Rm{S}}
\\
&- \Rm{v}_{\Rm{N}}^j(\bar{\uprho} + \bar{\Rm{p}} + \updelta \uprho_{\Rm{S}}) \partial_i (\Rm{B}_j + \Rm{A}_j)  -(\partial_i \Rm{U_S})\big( \updelta \uprho + \updelta \Rm{p} - 2(\Rm{U_S} + \upphi)(\bar{\uprho} + \bar{\Rm{p}} + \updelta \uprho_{\Rm{S}})  \big)
\\
 & - \updelta \uprho_\Rm{M} \partial_i (\upphi + {\Rm{U_M}}) - (\bar{\uprho} + \bar{\Rm{p}} + \updelta \uprho_{\Rm{S}})\partial_i\upphi_{\Rm{P}}  - \partial_i \updelta \Rm{p} \; ,
\end{aligned}
}
\end{equation}
where $\upphi_{\Rm{P}} = \upphi^{(0,4)} +  \upphi^{(1,2)}$.
This equation displays further interesting characteristics, for example, coupling between cosmological tensor and Newtonian vector and scalar perturbations. This can be seen in the term $\big(\Rm{h}_{ij} \Rm{v}_{\Rm{N}}^j (\bar{\uprho} + \bar{\Rm{p}} + \updelta \uprho_{\Rm{S}})\big)'$, and should be expected to result in new physical effects due to the interplay between perturbations on different length scales. This evolution equation was determined directly from the spatial components of the stress-energy conservation equation, and is required to consistently evolve the source terms in Eq. (\ref{rhoCcontinuity}), even though it does not appear in the field equations itself (at the order we are considering).

\section{Conclusions} \label{sec:conc}

We have derived and presented the relativistic Euler equations that exist in the two-parameters expansion proposed in Refs.  \cite{SRGTCKM} and \cite{Goldberg:2017}. These equations describe the evolution of density perturbations and peculiar velocities for a self-gravitating perfect fluid in an FLRW background. These equations are written down in gauge-invariant variables, and were used to confirm that the constraint equations from the two-parameter perturbation expansion are consistently evolved, despite the fact that terms can change size under differentiation. This gives confidence that the scheme is internally self-consistent and complete, and can be used to model the relativistic effects of non-linear structures in perturbation theory. 

The resulting Euler equations for the inhomogeneous part of the leading-order matter density and the peculiar velocity, together with the leading order gravitational Poisson equation, reproduce the standard results of Newtonian perturbation theory on an expanding background, as long as cosmological contributions to the peculiar velocity are included. These leading-order equations have well-known solutions in terms of Green's functions and numerical $N$-body simulations, as well as approximations that can be obtained by applying Eulerian perturbation theory \cite{Bernardeau, Goroff, Scoccimarro}. Subsequent higher-order equations that govern the leading-order contributions to the large-scale gravitational potentials are then given as linear partial differential equations that contain the known solutions to the lower-order Newtonian equations as source terms. In a sense, one can consider the equations for the cosmological quantities as being the result of performing a linear cosmological perturbation theory expansion on a background that is allowed to contain Newtonian gravitational fields (or vice versa). This explicitly shows the link between gravitational fields on large and short scales which occurs due to the non-linearity of Einstein's equations.

We expect that further investigation of the properties of this set of equations will lead to new insights about the nature of the interplay between short-scale non-linearity and large-scale fluctuations in the real Universe, and in particular that the relativistic Euler equations derived in this paper could be used to evolve matter fields both for the prediction of specific relativistic observables, and for inclusion in simulations. As was the case for the field equations presented in Section \ref{sec:2p}, the relativistic Euler equations presented above can be seen to contain mode-mixing terms between scalar, vector and tensor degrees of freedom. This does not occur at the same order in standard cosmological perturbation theory, and opens the possibility of experimental tests of gravity based on the production of what would normally be decoupled degrees of freedom.

Further directions for analytic exploration in this new formalism remain to be explored. These include the connections and differences with the post-Friedmann framework developed in \cite{bruni0}, the use of renormalised Eulerian perturbation theory to derive solutions to the leading-order system (see e.g. \cite{CrocceSocciRPT:2006}), and the extension of the perfect fluid treatment presented here to a full effective field theory treatment of the Boltzmann hierarchy (in analogy to Refs. \cite{Baumann:2010tm, EFTofLSS:2012}). We expect this to lead to a greater understanding of Einstein's theory and the effect of non-linear structure in the real Universe.

\section*{Acknowledgements}

\noindent
We are grateful to Karim Malik and Chris Clarkson  for helpful discussions. CSG and TC are supported by the STFC under grants ST/K50225X/1 and ST/N504257/1. The tensor algebra packages xAct \cite{xAct} and xPand \cite{xPand1, xPand2} were used in this work. 
\clearpage
\appendix

\section{Glossary} \label{glossary}

\subsection{Metric Perturbations}\label{MetricPerturbations}

The components of the metric perturbation tensor, $h_{\mu\nu}$, can be written in two parameter perturbation theory as

\begin{align}
h_{00} =& \; h_{00}^{(0,2)} + h_{00}^{(1,0)} + h_{00}^{(1,1)} + h_{00}^{(0,4)} + h_{00}^{(1,2)} + h_{00}^{(1,3)} + h_{00}^{(0,5)} + \dots \;, \\ 
h_{ij} =& \; h_{ij}^{(0,2)} + h_{ij}^{(1,0)} + h_{ij}^{(1,1)} + h_{ij}^{(0,4)} + h_{ij}^{(1,2)} + h_{ij}^{(1,3)} + h_{ij}^{(0,5)} + \dots \;,\\
h_{0i} =& \; h_{0i}^{(0,3)} + h_{0i}^{(1,0)} + h_{0i}^{(1,2)} + \dots \; ,
\end{align}
where each $Q^{(m,n)}$ is a metric perturbation, of the type used to construct gauge-invariant perturbations in \cite{Goldberg:2017} and \cite{SRGTCKM}. We have relabelled and scaled various different combinations of these perturbations throughout the paper, in order to bring the definitions into line with the literature on cosmological perturbation theory. In order to recap, here we provide a list of these different redefinitions:

\begin{align}
\Rm{U} =& \; -\frac{1}{2}  h_{00}^{(0,2)} + h_{00}^{(1,1)} = -\frac{1}{6} h_{ii}^{(0,2)} + h_{ii}^{(1,1)} \;,\\
\Rm{U}_{\Rm{S}} =& \; -\frac{1}{2}h_{00}^{(0,2)} =  -\frac{1}{6} h_{ii}^{(0,2)} \; ,  \\
\Rm{U}_{\Rm{M}} =& \; -\frac{1}{2} h_{00}^{(1,1)} = -\frac{1}{6} h_{ii}^{(1,1)} \;, \\
\upphi =& \; -\frac{1}{2} \Big( h_{00}^{(1,0)} + h_{00}^{(1,2)} + \frac{1}{2}h_{00}^{(0,4)} \Big) \;, \\
\uppsi =& \; -\frac{1}{2} \Big( h_{ii}^{(1,0)}  +  h_{ii}^{(1,2)} +  \frac{1}{2} h_{ii}^{(0,4)} \Big) \;, \\
\upphi_{5} =& \; -\frac{1}{2} \Big( h_{00}^{(1,3)}  +  h_{00}^{(0,5)}  \Big) \;,\\
\uppsi_{5} =& \; -\frac{1}{6} \Big( h_{ii}^{(1,3)}  +  h_{ii}^{(0,5)}  \Big) \;, \\
\Rm{S}_i =& \; - \big( h_{0i}^{(0,3)} + h_{0i}^{(1,0)} + h_{0i}^{(1,2)} \big) \;, \\
\Rm{A}_i =& \; -\big(  h_{0i}^{(1,0)} + h_{0i}^{(1,2)} \big) \;, \\
\Rm{B}_i =& \; -h_{0i}^{(0,3)} \;,\\
\upphi_{\Rm{P}} =& -\frac{1}{2}\Big(h_{00}^{(0,4)} +  h_{00}^{(1,2)}\Big) \; .
\end{align}

\subsection{Matter Perturbations}\label{MatterPerturbations}

The stress-energy tensor can be written
\begin{align}
T_{\mu\nu} = (\uprho + \Rm{p})u_{\mu}u_{\nu} + \Rm{p} g_{\mu\nu} \;,
\end{align}
where $u_{\mu} = (u_0,v_i)$, and $u_0$ can be determined in terms of metric perturbations using the four-vector normalisation condition. The two parameter expanded matter fields are then
\begin{align}
\uprho &= \uprho^{(0,0)} + \uprho^{(0,2)} + \uprho^{(1,0)} + \uprho^{(1,1)} + \uprho^{(1,2)} + \frac{1}{2} \uprho^{(0,4)}  \dots \; ,\\
\Rm{p} &= \Rm{p}^{(0,0)} + \Rm{p}^{(1,0)} + \Rm{p}^{(1,2)} + \frac{1}{2} \Rm{p}^{(0,4)} + \Rm{p}^{(1,3)} + \Rm{p}^{(0,5)} \dots \;, \\
v_i &= v_i^{(0,1)} + v_i^{(1,0)} + v_i^{(0,2)} + v_i^{(1,1)} + v_i^{(0,3)} \dots \; ,
\end{align}
where here each $Q^{(m,n)}$ is a matter perturbation of the type considered in \cite{Goldberg:2017} and \cite{SRGTCKM}. As detailed in \cite{Goldberg:2017}, both $\uprho^{(0,0)}$ and $\Rm{p}^{(0,0)}$ are necessarily homogeneous. On the other hand, $\uprho^{(0,2)}$ can be split into a homogeneous component and an inhomogeneous component, $\uprho^{(0,2)} = \bar{\uprho}^{(0,2)} + \updelta \uprho^{(0,2)}$. The matter perturbations used throught this paper can are then given by the following list: 
\begin{align}
\bar{\uprho} =&\; \uprho^{(0,0)} + \bar{\uprho}^{(0,2)} \;, \\
\updelta\uprho_{\Rm{N}} =&\; \updelta\uprho^{(0,2)} +  \uprho^{(1,1)} \;, \\
\updelta\uprho_{\Rm{S}} =&\; \updelta\uprho^{(0,2)} \;, \\
\updelta\uprho_{\Rm{M}} =&\; \uprho^{(1,1)} \;, \\
\updelta \uprho =&\;  \uprho^{(1,0)} + \uprho^{(1,2)} + \frac{1}{2} \uprho^{(0,4)} \;, \\
\bar{\Rm{p}} =&\; \Rm{p}^{(0,0)} \;, \\
\updelta{\Rm{p}} =&\; \Rm{p}^{(1,0)} + \Rm{p}^{(1,2)} + \frac{1}{2} \Rm{p}^{(0,4)} + \Rm{p}^{(1,3)} + \Rm{p}^{(0,5)} \;, \\
\Rm{v}_{\Rm{N}i} =&\; v_i^{(0,1)} \; ,\\
\Rm{v}_i =&\; v_i^{(1,0)} + v_i^{(0,2)} \;, \\
\Rm{v}_{\Rm{M}i} =&\; v_i^{(1,1)} +  v_i^{(0,3)} \;,\\
\Rm{v}_{\Rm{C}i} = &\; v_i^{(1,0)} \;,\\
\Rm{v}_{\Rm{P}i} = &\; v_i^{(0,2)} \;.
\end{align}

\section{Derivation of the conservation equations for cosmological perturbations} \label{appendix}

This appendix contains the derivation of the cosmological Euler equations that were withheld from Section \ref{sec:PerturbedConservationEquations}. These equations are found by differentiating the constraint equations (\ref{psieq}) and (\ref{seq}), and by using the evolution equations (\ref{phieq}) and (\ref{heq}).  As these equations contain post-Newtonian, mixed and cosmological terms, the reader must take into account the fact that spatial derivatives act differently on different types of perturbed quantities, as explained in Section \ref{sec:2p}. For example, carrying out calculations involving conformal time derivatives of the $\mathcal{O}(\eta^4)$ Einstein equations requires some knowledge of the $\mathcal{O}(\eta^5)$ equations. 

\subsection{Derivation of the $\mathcal{O}(\eta^5)$ continuity equation}

The conformal time derivative of Eq. (\ref{psieq}) is given by
\begin{align}
\frac{1}{3}\nabla\uppsi' - \mathcal{H}\uppsi'' - \mathcal{H}'\uppsi' - 2\mathcal{H}\mathcal{H}'\upphi - \mathcal{H}^2 \upphi' 
=\;\;&\frac{4\pi a^2}{3}\big( \updelta \uprho' + \updelta \uprho_{\rm eff}' +2\mathcal{H}(\updelta \uprho + \updelta \uprho_{\rm eff})\big) +\frac{1}{3}\mathcal{D}^{ij}\Rm{U}_{\Rm{S}}'\Rm{h}_{ij}\nonumber \\
&-\frac{16\pi a^2}{3}\big( \updelta \uprho_{\Rm{S}} \uppsi' + \uppsi(\updelta \uprho_{\Rm{S}}' + 2\mathcal{H}\updelta \uprho_{\Rm{S}})\big) + \frac{1}{3}\mathcal{D}^{ij}\Rm{U}_{\Rm{S}}\Rm{h}_{ij}' \;\;.
\end{align}
Substituting in for $\uppsi''$ using Eq. (\ref{phieq}) yields
\begin{align} \label{evolutionsubbedconstraintdash}
\nabla^2 (\uppsi' + \mathcal{H}\upphi)  =& - 3(\mathcal{H}^2 - \mathcal{H}')\uppsi' +4\pi a^2\big(\updelta \uprho' + 3\mathcal{H}(\updelta \uprho + \updelta \Rm{p})\big) 
+ 4\pi a^2\big(\updelta \uprho_{\rm eff}' + 3\mathcal{H}(\updelta \uprho_{\rm eff} + \updelta \Rm{p}_{\rm eff})\big)  + 8\pi a^2 \mathcal{H} \updelta \uprho_{\Rm{S}} \upphi  \nonumber \\
&+ \mathcal{D}^{ij}\Rm{U}_{\Rm{S}}'\Rm{h}_{ij} + \mathcal{D}^{ij}\Rm{U}_{\Rm{S}}\Rm{h}_{ij}' + 2\mathcal{H} \mathcal{D}^{ij}\Rm{U}_{\Rm{S}}\Rm{h}_{ij} 
- 16\pi a^2 \big( \updelta \uprho_{\Rm{S}} \uppsi' + \uppsi(\updelta \uprho_{\Rm{S}}' + 2\mathcal{H}\updelta \uprho_{\Rm{S}})\big) - 8\pi a^2 \mathcal{H} \updelta \uprho_{\Rm{S}} \uppsi\; .
\end{align}
On the left-hand side of this equation, $\nabla^2 (\uppsi' + \mathcal{H}\upphi) $ is obtained from the divergence of the vector Einstein equation at $\mathcal{O}(\eta^5)$. Expanding this equation gives
\begin{align} \label{divergenceofvectoreta5} 
\nabla^2(\uppsi' + \mathcal{H}\upphi) = &\;\;4\pi a^2 \partial^i \big(\Rm{v}_{\Rm{N}i}(\bar{\uprho} + \bar{\Rm{p}} + \updelta \uprho_{\Rm{S}} ) \Rm{U}_{\Rm{S}} \big) - 4\pi a^2 \partial^i\big(\Rm{v}_{\Rm{M}i}(\bar{\uprho} + \bar{\Rm{p}} +\updelta \uprho_{\Rm{S}})\big) - 4\pi a^2 \partial^i \big(\Rm{v}_{\Rm{N}i}(\bar{\uprho} + \bar{\Rm{p}} + \updelta \uprho_{\Rm{S}} ) \big) \upphi      \nonumber \\
 &+  8\pi a^2 \partial^i \big(\Rm{v}_{\Rm{N}i}(\bar{\uprho} + \bar{\Rm{p}} + \updelta \uprho_{\Rm{S}} ) \big) \uppsi 
 - 4\pi a^2 \big(\bar{\uprho} + \bar{\Rm{p}} + \updelta \uprho_{\Rm{S}} ) \partial^i\Rm{v}_{\Rm{C}i} - 4\pi a^2 \Rm{v}_{i} (\partial^i\updelta \uprho_{\Rm{M}}) - 4\pi a^2\partial^i \big( \Rm{v}_{\Rm{N}i} (\updelta \uprho + \updelta \Rm{p}) \big) \nonumber \\
 &+ 2\mathcal{H} \Rm{U}_{\Rm{S}} \nabla^2 \Rm{U}_{\Rm{S}} - \Rm{U}_{\Rm{S}}'\nabla^2\Rm{U}_{\Rm{S}} + 2\mathcal{H} \upphi \nabla^2 \Rm{U}_{\Rm{S}} - \uppsi' \nabla^2\Rm{U}_{\Rm{S}} - 2\Rm{U}_{\Rm{S}} \nabla^2 \Rm{U}_{\Rm{S}}' - 2\uppsi \nabla^2 \Rm{U}_{\Rm{S}}' + 2\mathcal{H} (\partial_i \Rm{U}_{\Rm{S}})(\partial^i \Rm{U}_{\Rm{S}}) \nonumber \\
 &- 3(\partial_i \Rm{U}_{\Rm{S}}')(\partial^i \Rm{U}_{\Rm{S}}) + \Rm{h}_{ij}' \mathcal{D}^{ij}\Rm{U}_{\Rm{S}} + \Rm{h}_{ij}\mathcal{D}^{ij}\Rm{U}_{\Rm{S}}' + 2\mathcal{H} \Rm{h}_{ij} \mathcal{D}^{ij}\Rm{U}_{\Rm{S}} - 2\pi a^2 \:\partial_i ( \rm{v}_{\rm{N}}^2 \:\rm{v}_{\rm{N}}^{\it{i}}(\bar{\uprho} + \bar{\rm{p}} +\updelta\uprho_{\rm{S}}) ) \nonumber \\
  &- 4\pi a^2 \partial^{\It{i}} \Rm{v}_{\Rm{P}\It{i}} \updelta \uprho_{\Rm{M}}\, .
\end{align}
Substituting (\ref{divergenceofvectoreta5}) into (\ref{evolutionsubbedconstraintdash}), cancelling tensorial terms and simplifying using the lower-order conservation equations, we obtain 
\begin{align} \label{midpoint}
\updelta \uprho' + 3\mathcal{H}(\updelta \uprho + \updelta \Rm{p}) 
=&- \updelta \uprho_{\rm eff}' - 3\mathcal{H}(\updelta \uprho_{\rm eff} + \updelta \Rm{p}_{\rm eff}) +\partial^i\big( \Rm{v}_{\Rm{N}i}(\bar{\uprho} + \bar{\Rm{p}} + \updelta \uprho_{\Rm{S}})\Rm{U}_{\Rm{S}}\big) -  \upphi \partial^i\big(\Rm{v}_{\Rm{N}i}(\bar{\uprho} + \bar{\Rm{p}} + \updelta \uprho_{\Rm{S}})\big) \nonumber \\
&-  (\bar{\uprho} + \bar{\Rm{p}} + \updelta \uprho_{\Rm{S}})\partial^i \Rm{v}_{\Rm{C}i}
 - \partial^i\big(\Rm{v}_{\Rm{N}i} (\updelta \uprho + \updelta \Rm{p} )\big)  + 3 (\bar{\uprho} + \bar{\Rm{p}} + \updelta \uprho_{\Rm{S}}) \uppsi'  - \partial^i\big(\Rm{v}_{\Rm{M}i}(\bar{\uprho} + \bar{\Rm{p}} +\updelta \uprho_{\Rm{S}})\big) \nonumber \\
&  + \frac{1}{4\pi a^2}\Big(2\mathcal{H} \Rm{U}_{\Rm{S}} \nabla^2 \Rm{U}_{\Rm{S}} - \Rm{U}_{\Rm{S}}' \nabla^2 \Rm{U}_{\Rm{S}} - 2\Rm{U}_{\Rm{S}} \nabla^2 \Rm{U}_{\Rm{S}}' + 2\mathcal{H} (\partial_i \Rm{U}_{\Rm{S}})(\partial^i \Rm{U}_{\Rm{S}})- 3 (\partial_i \Rm{U}_{\Rm{S}}')(\partial^i \Rm{U}_{\Rm{S}}) \Big)\nonumber \\ 
&- \frac{1}{2} \:\partial_i ( \rm{v}_{\rm{N}}^2 \:\rm{v}_{\rm{N}}^{\it{i}}(\bar{\uprho} + \bar{\rm{p}} +\updelta\uprho_{\rm{S}}) ) - \Rm{v}_{\It{i}} (\partial^{\It{i}} \updelta \uprho_{\Rm{M}}) - \partial^{\It{i}} \Rm{v}_{\Rm{P}\It{i}} \updelta \uprho_{\Rm{M}}  \;   .
\end{align}
At this point it is useful to note the following relations, which can be derived without difficulty from lower-order conservation equations:
\begin{align} 
 \partial^i\big( \Rm{v}_{\Rm{N}i}(\bar{\uprho} + \bar{\Rm{p}} + \updelta \uprho_{\Rm{S}})\Rm{U}_{\Rm{S}}\big)  &= ( \partial^i \Rm{U}_{\Rm{S}})\Rm{v}_{\Rm{N}i}(\bar{\uprho} + \bar{\Rm{p}} + \updelta \uprho_{\Rm{S}}) - \frac{1}{4\pi a^2} (\Rm{U}_{\Rm{S}}\nabla^2 \Rm{U}_{\Rm{S}}' + \mathcal{H} \Rm{U}_{\Rm{S}} \nabla^2 \Rm{U}_{\Rm{S}}) \;, \\
3\Rm{U}_{\Rm{S}}' (\updelta \uprho_{\Rm{N}} + \bar{\uprho} + \bar{\Rm{p}}) &=- \frac{1}{4\pi a^2}(-3\Rm{U}_{\Rm{S}}' \nabla^2 \Rm{U}_{\Rm{S}} - 3 \Rm{U}_{\Rm{S}}' \mathcal{H}^2 + 3\Rm{U}_{\Rm{S}}' \mathcal{H}') \; , \\
\frac{1}{4\pi a^2}\Rm{U}_{\Rm{S}} \nabla^2 (\Rm{U}_{\Rm{S}}' + \mathcal{H} \Rm{U}_{\Rm{S}}) &= - \Rm{U}_{\Rm{S}} \;\partial^i\big(\Rm{v}_{\Rm{N}i}(\bar{\uprho} + \bar{\Rm{p}} + \updelta \uprho_{\Rm{S}})\big)\;. 
\end{align}
These relations can be exploited to manipulate Eq. (\ref{midpoint}) into the following form:
\begin{align}
&\updelta \uprho' + 3\mathcal{H}(\updelta \uprho + \updelta \Rm{p}) + \big(\updelta \uprho_{\rm eff}' + 3\mathcal{H}(\updelta \uprho_{\rm eff} + \updelta \Rm{p}_{\rm eff})\big) \nonumber \\
=\;\;\;& (\partial^i\Rm{U}_{\Rm{S}}) \Rm{v}_{\Rm{N}i}(\bar{\uprho} + \bar{\Rm{p}} + \updelta \uprho_{\Rm{S}}) -  \upphi \partial^i\big( \Rm{v}_{\Rm{N}i}(\bar{\uprho} + \bar{\Rm{p}} + \updelta \uprho_{\Rm{S}})\big) - \partial^i\big(\Rm{v}_{\Rm{M}i}(\bar{\uprho} + \bar{\Rm{p}} +\updelta \uprho_{\Rm{S}})\big)  \nonumber \\
&- \Rm{v}_{i} (\partial^i\updelta \uprho_{\Rm{M}})- \partial^i\Rm{v}_{\Rm{P}i} \updelta \uprho_{\Rm{M}}   -  (\bar{\uprho} + \bar{\Rm{p}} + \updelta \uprho_{\Rm{S}})\partial^i \Rm{v}_{\Rm{C}i} + 3\Rm{U}_{\Rm{S}}' (\bar{\uprho} + \bar{\Rm{p}} + \updelta \uprho_{\Rm{S}} ) + 3 (\bar{\uprho} + \bar{\Rm{p}} + \updelta \uprho_{\Rm{S}}) \uppsi'   \nonumber \\
& - \partial^i \big( \Rm{v}_{\Rm{N}i} (\updelta \uprho + \updelta \Rm{p} )\big) - \Rm{U}_{\Rm{S}} \partial^i\big(\Rm{v}_{\Rm{N}i}(\bar{\uprho} + \bar{\Rm{p}} + \updelta \uprho_{\Rm{S}})\big) - \frac{1}{2} \:\partial_i ( \rm{v}_{\rm{N}}^2 \:\rm{v}_{\rm{N}}^{\it{i}}(\bar{\uprho} + \bar{\rm{p}} +\updelta\uprho_{\rm{S}}) ) \nonumber \\
&  + \frac{1}{4\pi a^2}\Big(- 4\Rm{U}_{\Rm{S}}' \nabla^2 \Rm{U}_{\Rm{S}} - 3\Rm{U}_{\Rm{S}}'\mathcal{H}^2 + 3 \Rm{U}_{\Rm{S}}' \mathcal{H}' - 4\Rm{U}_{\Rm{S}} \nabla^2 \Rm{U}_{\Rm{S}}' + 2\mathcal{H}(\partial_i \Rm{U}_{\Rm{S}})(\partial^i \Rm{U}_{\Rm{S}})  - 3 (\partial_i \Rm{U}_{\Rm{S}}')(\partial^i \Rm{U}_{\Rm{S}})  \Big)  \;.
\end{align}

Now, by considering the following combination of effective fluid quantities,
\begin{align}
\updelta \uprho_{\rm eff} ' \;+\; &3\mathcal{H}(\updelta \uprho_{\rm eff} + \updelta \Rm{p}_{\rm eff} ) = \big( (\bar{\uprho} + \bar{\Rm{p}} + \updelta \uprho_{\Rm{S}})\Rm{v}_{\Rm{N}}^2\big)' + 4\mathcal{H} \Rm{v}_{\Rm{N}}^2(\bar{\uprho} + \bar{\Rm{p}} + \updelta \uprho_{\Rm{S}})  \nonumber \\
+ &\frac{1}{4\pi a^2} \Big( -3\mathcal{H}^2 \Rm{U}_{\Rm{S}}' + 3\mathcal{H}'\Rm{U}_{\Rm{S}}' - 4\Rm{U}_{\Rm{S}}' \nabla^2\Rm{U}_{\Rm{S}} - 4\Rm{U}_{\Rm{S}}\nabla^2 \Rm{U}_{\Rm{S}}' + 2\mathcal{H}(\partial_i \Rm{U}_{\Rm{S}})(\partial^i \Rm{U}_{\Rm{S}})  - 3 (\partial_i \Rm{U}_{\Rm{S}}')(\partial^i \Rm{U}_{\Rm{S}})  \Big)\;,
\end{align}
it is easy to see that we have obtained precisely Eq. (\ref{rhoCcontinuity}). This equation has been checked by comparing with the time component of the $\mathcal{O}(\eta^5)$ stress-energy conservation equation.

\subsection{Derivation of the $\mathcal{O}(\eta^5)$ Euler equation}

The conformal time derivative of Eq. (\ref{seq}) is given by
\begin{align} \label{Adash}
0=& \nabla^2 \Rm{A}'_i   + 4\partial_i \big( \uppsi''  + \mathcal{H}'\upphi + \mathcal{H}\upphi' \big) 
+16\pi a^2 \Big( (\bar{\uprho}+ \bar{\Rm{p}} + \updelta \uprho_{\Rm{S}})'( \Rm{v}_{i} - \Rm{A}_i) + (\bar{\uprho}+ \bar{\Rm{p}} + \updelta \uprho_{\Rm{S}})( \Rm{v}_{i} - \Rm{A}_i)' + \Rm{Q}_i^{\prime \;\;\rm eff }\Big)  \nonumber \\
&+16\pi a^2 \Big( 2\mathcal{H}(\bar{\uprho}+ \bar{\Rm{p}} + \updelta \uprho_{\Rm{S}})( \Rm{v}_{i} - \Rm{A}_i) + 2\mathcal{H} \Rm{Q}_i^{\prime \;\; \rm eff} \Big)
+8\pi a^2 \updelta \uprho'_{\Rm{S}} \Rm{A}_i + 8\pi a^2 \updelta \uprho_{\Rm{S}} \Rm{A}'_i + 16\mathcal{H}\pi a^2 \updelta \uprho_{\Rm{S}}'\Rm{A}_i \nonumber \\
& + 2\Rm{A}^{j \prime} \partial_i \partial_j\Rm{U}_{\Rm{S}}  + 2\Rm{A}^j\partial_i\partial_j\Rm{U}_{\Rm{S}}'  \;.
\end{align}
In order to proceed further, it is now necessary to eliminate the term $\nabla^2 \Rm{A}_i' $. We achieve this by using the divergence of the trace-free part of the $ij$-field equation, $\partial^jG_{\langle ij \rangle} - 8\pi \partial^jT_{\langle ij \rangle} =0$. Due to the length of the expressions that result, we choose to split the calculation into three sections; the divergence of the trace-free field equation at $\mathcal{O}(\eta^5)$, the $\mathcal{O}(\eta^5)$ gradients of the scalar field equations, and the conservation equation itself. 

\subsubsection{Divergence of $\mathcal{O}(\eta^5)$  trace-free field equation}

The $\mathcal{O}(\eta^5)$ component of the divergence of the trace-free $ij$ field equation is given by
\begin{align} \label{divij}
\nabla^2 \Rm{A}_i' =\;\;\; & 16\pi a^2(\Rm{v}_{i} - \Rm{A}_i) \partial^j \big((\bar{\uprho} + \bar{\Rm{p}} + \updelta \uprho_{\Rm{S}})\Rm{v}_{\Rm{N}j}\big) +16\pi a^2\partial^j \Rm{v}_{\Rm{P}i}\big((\bar{\uprho} + \bar{\Rm{p}} + \updelta \uprho_{\Rm{S}})\Rm{v}_{\Rm{N}j}\big) \nonumber \\
& + 16\pi a^2(\Rm{v}_j - \Rm{A}_j) \partial^j \big((\bar{\uprho} + \bar{\Rm{p}} + \updelta \uprho_{\Rm{S}})\Rm{v}_{\Rm{N}i}\big)  + 16\pi a^2\partial^j \Rm{v}_{\Rm{P}j} \big((\bar{\uprho} + \bar{\Rm{p}} + \updelta \uprho_{\Rm{S}})\Rm{v}_{\Rm{N}i}\big) \nonumber \\
 & -\frac{32\pi a^2}{3}(\Rm{v}^j - \Rm{A}^j)\partial_i \big((\bar{\uprho} + \bar{\Rm{p}} + \updelta \uprho_{\Rm{S}})\Rm{v}_{\Rm{N}j}\big)  -\frac{32\pi a^2}{3}\partial_i \Rm{v}_{\Rm{P}}^{j} \big((\bar{\uprho} + \bar{\Rm{p}} + \updelta \uprho_{\Rm{S}})\Rm{v}_{\Rm{N}j}\big) \nonumber \\
&+ 8\mathcal{H} \partial_i(\uppsi' + \mathcal{H}\upphi) + 32\mathcal{H} \pi a^2 \big((\bar{\uprho} + \bar{\Rm{p}} + \updelta \uprho_{\Rm{S}})(\Rm{v}_{i} - \Rm{A}_i) + \Rm{Q}_i^{\Rm{eff}}\big)  \nonumber \\
& + \frac{8\mathcal{H}}{3} \Rm{A}^j \partial_j \partial_i \Rm{U}_{\Rm{S}} - \frac{2}{3} \Rm{A}'^j \partial_j \partial_i \Rm{U}_{\Rm S} - \frac{2}{3} \Rm{A}^j \partial_j\partial_i \Rm{U}_{\Rm S}'  -\frac{4}{3} \partial^k \partial^j \Rm{U}_{\Rm S} \partial_i \Rm{h}_{kj} - \frac{4}{3} \Rm{h}^{kj} \partial_k \partial_j \partial_i \Rm{U}_{\Rm{M}}\nonumber \\
&-16\pi a^2  \mathcal{H} \updelta \uprho_{\Rm{S}} \Rm{A}_i -8\pi a^2  \updelta \uprho_{\Rm{S}} \Rm{A}'_i -8\pi a^2   \updelta \uprho'_{\Rm{S}} \Rm{A}_i - \Rm{A}^j \partial_j \nabla^2 \Rm{B}_i + \frac{4}{3} \Rm{A}^j \partial_i \nabla^2 \Rm{B}_j \nonumber \\
& -\frac{28}{3}\partial_j \partial_i \Rm{U}_{\Rm{M}} \partial^j\Rm{U}_{\Rm S}  -\frac{28}{3}\partial_j \partial_i \Rm{U}_{\Rm{S}} \partial^j\Rm{U}_{\Rm{M}} - 4 \partial_j \partial_i \Rm{U}_{\Rm S}  \partial^j \upphi - \frac{16}{3} \partial_j \partial_i \Rm{U}_{\Rm S}  \partial^j \uppsi \nonumber \\
&-\frac{64\pi a^2}{3} \Rm{U}_{\Rm{M}} \partial_i \updelta \uprho_{\Rm{S}} -\frac{64\pi a^2}{3} \Rm{U}_{\Rm S} \partial_i \updelta \uprho_{\Rm{M}} -\frac{32\pi a^2}{3} \upphi \partial_i \updelta \uprho_{\Rm{M}} -\frac{32\pi a^2}{3} \uppsi \partial_i \updelta \uprho_{\Rm{M}} \nonumber \\
&-\frac{16\pi a^2}{3} \updelta \uprho_{\Rm{M}} \partial_i \Rm{U}_{\Rm S} -\frac{16\pi a^2}{3}  \updelta \uprho_{\Rm{S}} \partial_i \Rm{U}_{\Rm{M}} -\frac{32\pi a^2}{3} \updelta \uprho_{\Rm{S}} \partial_i \uppsi + \frac{16\pi a^2}{3} \updelta \uprho_{\Rm{S}} \partial_i \upphi \nonumber \\
&+16 \pi a^2 \partial^j(\Rm{v}_{\Rm{N}\It{i}}\Rm{v}_{\Rm{N}\It{j}}\updelta \uprho_{\Rm M}) -\frac{16 \pi a^2}{3}\partial_i(\Rm{v}_{\Rm N}^2\updelta \uprho_{\Rm M}) - \frac{4}{3}\partial_i\nabla^2(\uppsi + \uppsi_{5} - \upphi - \upphi_{5})\;.
\end{align}
It should be noted that the term $ \frac{4}{3}\partial_i\nabla^2(\uppsi + \uppsi_{5} - \upphi - \upphi_{5})$ involves the higher-order potentials $\upphi_{5} \equiv -\frac{1}{2}( h^{(0,5)} + h^{(1,3)} )$ and $\uppsi_{5} \equiv -\frac{1}{2}( h^{(0,5)} + h^{(1,3)} )$. These terms do not appear in the scalar field equations at $\mathcal{O}(\eta^4)$, however, they do appear in the $\mathcal{O}(\eta^5)$ field equations, and so cannot be neglected in this calculation. Fortunately, these terms can also be obtained in terms of products of lower-order quantities by calculating gradients of the $\mathcal{O}(\eta^5)$ field equations, enabling us to eliminate them in favour of quantities whose evolution is already known.

\subsubsection{Gradients of the $\mathcal{O}(\eta^5)$ scalar equations}

The relevant linear combination of gradients of the $\mathcal{O}(\eta^5)$ scalar field equations that allows us to eliminate the higher-order potentials from the $\mathcal{O}(\eta^5)$ part of the divergence of the trace-free $ij$-field equation is
\begin{align} \label{gradscalar}
&\frac{4}{3}\partial_i\nabla^2(\uppsi + \uppsi_{5} - \upphi - \upphi_{5})  \nonumber \\
 =&\frac{8\mathcal{H}}{3} \Rm{A}^j \partial_j \partial_i \Rm{U}_{\Rm S} + \frac{4}{3} \Rm{A}'^j \partial_j \partial_i \Rm{U}_{\Rm S} + \frac{4}{3} \Rm{A}^j \partial_j\partial_i \Rm{U}_{\Rm S}' -\frac{28}{3}\partial_j \partial_i \Rm{U}_{\Rm{M}} \partial^j \Rm{U}_{\Rm S}  -\frac{28}{3}\partial_j \partial_i \Rm{U}_{\Rm S} \partial^j \Rm{U}_{\Rm{M}} \nonumber \\
&  -\frac{4}{3} \partial^k \partial^j \Rm{U}_{\Rm S} \partial_i \Rm{h}_{kj} - \frac{4}{3} \Rm{h}^{kj} \partial_k \partial_j \partial_i \Rm{U}_{\Rm{M}}- 4 \partial_j \partial_i \Rm{U}_{\Rm S}  \partial^j \upphi - \frac{16}{3} \partial_j \partial_i \Rm{U}_{\Rm S}  \partial^j \uppsi -16\pi a^2 \partial_i\updelta \Rm{p} \nonumber \\
&  - \frac{16\pi a^2}{3}(\Rm{A}^j + 2\Rm{v}^j) \partial_i \big((\bar{\uprho} + \bar{\Rm{p}} + \updelta \uprho_{\Rm{S}})\Rm{v}_{\Rm{N}j}\big) - \frac{32\pi a^2}{3} \partial_i  \Rm{v}_{\Rm{P}}^j \big((\bar{\uprho} + \bar{\Rm{p}} + \updelta \uprho_{\Rm{S}})\Rm{v}_{\Rm{N}j}\big)\nonumber \\
&+\frac{1}{3} \Rm{A}^j \partial_i \nabla^2 \Rm{B}_j - 16\pi a^2 \updelta \uprho_{\Rm{M}} \partial_i \Rm{U}_{\Rm S} - 16\pi a^2 \updelta \uprho_{\Rm{S}} \partial_i \Rm{U}_{\Rm{M}} -\frac{64\pi a^2}{3} \Rm{U}_{\Rm{M}} \partial_i \updelta \uprho_{\Rm{S}} -\frac{64\pi a^2}{3} \Rm{U}_{\Rm S} \partial_i \updelta \uprho_{\Rm{M}} -\frac{16 \pi a^2}{3}\partial_i(\Rm{v}_{\Rm N}^2\updelta \uprho_{\Rm M}) \nonumber \\
& -\frac{32\pi a^2}{3} \upphi \partial_i \updelta \uprho_{\Rm{M}} -\frac{32\pi a^2}{3} \uppsi \partial_i \updelta \uprho_{\Rm{M}} 
-\frac{16\pi a^2}{3} \updelta \uprho_{\Rm{M}} \partial_i \Rm{U}_{\Rm S} -\frac{16\pi a^2}{3}  \updelta \uprho_{\Rm{S}} \partial_i \Rm{U}_{\Rm{M}} -\frac{32\pi a^2}{3} \updelta \uprho_{\Rm{S}} \partial_i \uppsi - \frac{32\pi a^2}{3} \updelta \uprho_{\Rm{S}} \partial_i \upphi \nonumber \\
&+ 4 \partial_i(\Rm{U}_{\Rm{M}}'' + 3\mathcal{H}\Rm{U}_{\Rm{M}}' + 2\mathcal{H}^2 \Rm{U}_{\Rm{M}} + \mathcal{H}'\Rm{U}_{\Rm{M}}) - 16\pi a^2(\bar{\uprho} +\bar{\Rm{p}}) \partial_i \Rm{U}_{\Rm{M}} + \partial_i (4 \mathcal{H} \upphi' + 2\mathcal{H}' \upphi + 4\mathcal{H}^2 \upphi) \nonumber \\
&+ 4 \partial_i(\uppsi'' + 2\mathcal{H}\uppsi' + \mathcal{H}^2 \uppsi + \frac{1}{2}\mathcal{H}'\uppsi) - 8\pi a^2(\bar{\uprho} -\bar{\Rm{p}})\partial_i \uppsi - 2\Lambda a^2 \partial_i \uppsi  - 8\pi a^2 (\bar{\uprho} + 3\bar{\Rm{p}}) \partial_i \upphi + 2\Lambda a^2 \partial_i \upphi \; ,
\end{align}
We now have all the results we need to demonstrate the conservation of the vector constraint equation.

\subsubsection{Vector conservation equation at $\mathcal{O}(\eta^5)$}

Using the result above, we can use Eq. (\ref{divij}) to substitute in for $\nabla^2 \Rm{A}'_i$ in Eq. (\ref{Adash}). This gives

\begin{align} \label{conservationeuler5part2}
&16 \pi a^2 \Big( (\bar{\uprho}+ \bar{\Rm{p}} + \updelta \uprho_{\Rm{S}})'( \Rm{v}_{i} - \Rm{A}_i) + (\bar{\uprho}+ \bar{\Rm{p}} + \updelta \uprho_{\Rm{S}})( \Rm{v}_{i} - \Rm{A}_i)' + \Rm{Q}_i^{\prime\;\rm eff } + 4\mathcal{H}(\bar{\uprho}+ \bar{\Rm{p}} + \updelta \uprho_{\Rm{S}})( \Rm{v}_{i} - \Rm{A}_i) + 4\mathcal{H} \Rm{Q}_i^{ \rm eff}      \nonumber \\
&+ (\Rm{v}_{i} - \Rm{A}_i) \partial^j \big((\bar{\uprho} + \bar{\Rm{p}} + \updelta \uprho_{\Rm{S}})\Rm{v}_{\Rm{N}j}\big) +\partial^j \Rm{v}_{\Rm{P}i}  \big((\bar{\uprho} + \bar{\Rm{p}} + \updelta \uprho_{\Rm{S}})\Rm{v}_{\Rm{N}j}\big) + (\Rm{v}_{j} - \Rm{A}_j) \partial^j \big((\bar{\uprho} + \bar{\Rm{p}} + \updelta \uprho_{\Rm{S}})\Rm{v}_{\Rm{N}i}\big)  \nonumber \\
&+ \partial^j \Rm{v}_{\Rm{P}j}  \big((\bar{\uprho} + \bar{\Rm{p}} + \updelta \uprho_{\Rm{S}})\Rm{v}_{\Rm{N}i}\big) \Big) +16 \pi a^2 \partial^j(\Rm{v}_{\Rm{N}\It{i}}\Rm{v}_{\Rm{N}\It{j}}\updelta \uprho_{\Rm M}) - \frac{32 \pi a^2}{3} \partial_i\Rm{v}_{\Rm{P}}^j \big( (\bar{\uprho} + \bar{\Rm{p}} + \updelta \uprho_{\Rm{S}})\Rm{v}_{\Rm{N}j} \big)\nonumber \\
& - \frac{32 \pi a^2}{3}(\Rm{v}^j - \Rm{A}^j)\partial_i \big( (\bar{\uprho} + \bar{\Rm{p}} + \updelta \uprho_{\Rm{S}})\Rm{v}_{\Rm{N}j} \big)  -\frac{16 \pi a^2}{3}\partial_i(\Rm{v}_{\Rm N}^2\updelta \uprho_{\Rm M})- \Rm{A}^j \partial_j \nabla^2 \Rm{B}_i + \frac{4}{3} \Rm{A}^j \partial_i \nabla^2 \Rm{B}_j   +\frac{8\mathcal{H}}{3} \Rm{A}^j \partial_j \partial_i \Rm{U}_{\Rm S}  \nonumber \\
&+ \frac{4}{3} \Rm{A}'^j \partial_j \partial_i \Rm{U}_{\Rm S} + \frac{4}{3} \Rm{A}^j \partial_j\partial_i \Rm{U}_{\Rm S}' -\frac{4}{3} \partial^k \partial^j \Rm{U}_{\Rm S} \partial_i \Rm{h}_{kj} - \frac{4}{3} \Rm{h}^{kj} \partial_k \partial_j \partial_i \Rm{U}_{\Rm{M}}  -\frac{28}{3}\partial_j \partial_i \Rm{U}_{\Rm{M}} \partial^j \Rm{U}_{\Rm S}  -\frac{28}{3}\partial_j \partial_i \Rm{U}_{\Rm S} \partial^j\Rm{U}_{\Rm{M}} \nonumber \\
&-\frac{64\pi a^2}{3} \Rm{U}_{\Rm{M}} \partial_i \updelta \uprho_{\Rm{S}} -\frac{64\pi a^2}{3} \Rm{U}_{\Rm S} \partial_i \updelta \uprho_{\Rm{M}} -\frac{32\pi a^2}{3} \upphi \partial_i \updelta \uprho_{\Rm{M}} -\frac{32\pi a^2}{3} \uppsi \partial_i \updelta \uprho_{\Rm{M}} -\frac{16\pi a^2}{3} \updelta \uprho_{\Rm{M}} \partial_i \Rm{U}_{\Rm S} -\frac{16\pi a^2}{3}  \updelta \uprho_{\Rm{S}} \partial_i \Rm{U}_{\Rm{M}}  \nonumber \\
&  -\frac{32\pi a^2}{3} \updelta \uprho_{\Rm{S}} \partial_i \uppsi + \frac{16\pi a^2}{3} \updelta \uprho_{\Rm{S}} \partial_i \upphi  + 4 \partial_i \uppsi'' + 8\mathcal{H} \partial_i \uppsi' + 4\mathcal{H}' \partial_i \uppsi' + 8\mathcal{H}^2 \partial_i \upphi  - \frac{4}{3}\partial_i\nabla^2(\uppsi + \uppsi_{5} - \upphi - \upphi_{5}) =0 \; . 
\end{align}
We can now substitute in for the gradients of the $\mathcal{O}(\eta^5)$ field equation, $\frac{4}{3}\partial_i\nabla^2(\uppsi + \uppsi_{5} - \upphi - \upphi_{5}) $, using Eq. (\ref{gradscalar}). After some manipulation we then obtain
\begin{align}
0 =&16 \pi a^2 \Big( (\bar{\uprho}+ \bar{\Rm{p}} + \updelta \uprho_{\Rm{S}})'( \Rm{v}_{i} - \Rm{A}_i) + (\bar{\uprho}+ \bar{\Rm{p}} + \updelta \uprho_{\Rm{S}})( \Rm{v}_{i} - \Rm{A}_i)' + \Rm{Q}_i^{\prime\;\rm eff }   + 4\mathcal{H}(\bar{\uprho}+ \bar{\Rm{p}} + \updelta \uprho_{\Rm{S}})( \Rm{v}_{i} - \Rm{A}_i) + 4\mathcal{H} \Rm{Q}_i^{\rm eff}    \nonumber \\
&+ \partial^j(\Rm{v}_{\Rm{N}\It{i}}\Rm{v}_{\Rm{N}\It{j}}\updelta \uprho_{\Rm M}) + (\Rm{v}_{i} - \Rm{A}_i) \partial^j \big((\bar{\uprho} + \bar{\Rm p} + \updelta \uprho_{\Rm S})\Rm{v}_{\Rm{N}j}\big) + (\Rm{v}_{j} - \Rm{A}_j) \partial^j \big((\bar{\uprho} + \bar{\Rm{p}} + \updelta \uprho_{\Rm{S}})\Rm{v}_{\Rm{N}i}\big) + \Rm{A}^j \partial_i \big( (\bar{\uprho} + \bar{\Rm{p}} + \updelta \uprho_{\Rm{S}})\Rm{v}_{\Rm{N}j}  \big)  \nonumber \\
&+ (\partial^j \Rm{v}_{\Rm{P} i})\Rm{v}_{\Rm{N}j}(\bar{\uprho} + \bar{\Rm{p}} + \updelta \uprho_{\Rm{S}}) +  (\partial^j \Rm{v}_{\Rm{P} j})\Rm{v}_{\Rm{N}i}(\bar{\uprho} + \bar{\Rm{p}} + \updelta \uprho_{\Rm{S}}) \Big) \nonumber \\
& - \Rm{A}^j \partial_j \nabla^2 \Rm{B}_i + \Rm{A}^j \partial_i \nabla^2 \Rm{B}_j + 16 \pi a^2 \partial_i \updelta \Rm{p}  + 16\pi a^2 \big((\bar{\uprho} + \bar{\Rm{p}} +  \updelta \uprho_{\Rm{S}}) \partial_i \Rm{U}_{\Rm{M}} + \updelta \uprho_{\Rm{M}} \partial_i \Rm{U}_{\Rm S} + \updelta \uprho_{\Rm{S}} \partial_i \upphi \big) \nonumber \\
& + 2\mathcal{H}' \partial_i \upphi + 4\mathcal{H}^2 \partial_i \upphi + 8\pi a^2 (\bar{\uprho} + 3\bar{\Rm{p}}) \partial_i \upphi - 2\Lambda a^2 \partial_i \upphi \nonumber \ - 2\mathcal{H}' \partial_i \uppsi - 4\mathcal{H}^2 \partial_i \uppsi + 8\pi a^2 (\bar{\uprho} - \bar{\Rm{p}}) \partial_i \uppsi + 2\Lambda a^2 \partial_i \uppsi \nonumber \\
&-4 \partial_i \Rm{U}''_{\Rm{M}} - 12 \mathcal{H} \partial_i \Rm{U}'_{\Rm{M}} -4 \mathcal{H}' \partial_i \Rm{U}_{\Rm{M}} - 8\mathcal{H}^2 \partial_i \Rm{U}_{\Rm{M}} \;.
\end{align}
Now using the following three relations:
\begin{align}
\Rm{Q}_i^{\prime \; \rm eff} + 4\mathcal{H} \Rm{Q}_i^{\rm eff} &= \updelta \uprho_{\Rm{M}}' \Rm{v}_{\Rm{N}i} + \updelta \uprho_{\Rm{M}} \Rm{v}_{\Rm{N}i}' + 4\mathcal{H}\updelta \uprho_{\Rm{M}} \Rm{v}_{\Rm{N}i} + \frac{1}{4\pi a^2}\partial_i \big( \Rm{U}''_{\Rm{M}} + 3\mathcal{H}'\Rm{U}_{\Rm{M}}' + (2\mathcal{H}^2 + \mathcal{H}')\Rm{U}_{\Rm{M}} \big) \\
\Rm{A}^j \partial_i \nabla^2 \Rm{B}_j- \Rm{A}^j \partial_j \nabla^2 \Rm{B}_i &= 16\pi a^2 \Rm{A}^j \Big(\partial_j \big((\bar{\uprho} + \bar{\Rm{p}} + \updelta \uprho_{\Rm{S}} \big) \Rm{v}_{\Rm{N}i}  - \partial_i \big((\bar{\uprho} + \bar{\Rm{p}} + \updelta \uprho_{\Rm{S}} \big) \Rm{v}_{\Rm{N}j} \Big)\\
2\mathcal{H}^2 + \mathcal{H}' &= 12 \pi a^2 (\bar{\uprho} + \bar{\Rm{p}})+ \Lambda a^2 \; ,
\end{align}
we can simplify to obtain Eq. (\ref{vCeuler}), which we confirm can also be directly derived from the stress-energy conservation equations. This calculation demonstrates that Eq. (\ref{seq}) is maintained under evolution, as required.

\end{widetext}

\end{document}